\documentclass[longauth]{aa}  
\usepackage{graphicx}
\usepackage{txfonts}
\usepackage{txfonts}
\usepackage{booktabs}
\usepackage{siunitx}
\usepackage{amsmath}
\usepackage{threeparttable}
\usepackage{cuted}

\usepackage{dsfont}

\newcommand{\aaj}{\emph{Astronomy \& Astrophysics}}

\newcommand{\sci}{\emph{Science}}

\usepackage{tikz}
\usetikzlibrary{arrows.meta}
\usetikzlibrary{shapes.arrows}
\usetikzlibrary{calc}
\usepackage{siunitx}
\usepackage{subcaption}
\usepackage{pgfplots}
\pgfplotsset{compat=1.7}
\usepackage{ulem}
\usepackage[etex=true,export]{adjustbox}
\usepackage[version=4]{mhchem}
\usepackage{orcidlink}

\makeatletter
\newcommand\notsotiny{\@setfontsize\notsotiny{7.5}{7}}
\makeatother

\usepackage{tabularx}
    \newcolumntype{L}{>{\raggedright\arraybackslash}X}

\begin{document} 

   \title{Panchromatic characterization of the Y0 brown dwarf WISEP~J173835.52+273258.9 using JWST/MIRI}
\author{
M.~Vasist\orcidlink{0000-0002-4511-3602}\inst{\ref{star_liege},\ref{mont_liege}}
\and P.~Molli\`ere\orcidlink{0000-0003-4096-7067}\inst{\ref{mpia}} 
\and H.~K\"uhnle\inst{\ref{eth}} 
\and O.~Absil\orcidlink{0000-0002-4006-6237}\inst{\ref{star_liege}} 
\and G.~Louppe\orcidlink{0000-0002-2082-3106}\inst{\ref{mont_liege}}
\and Rens Waters\orcidlink{0000-0002-5462-9387}\inst{\ref{radboud},\ref{hfml},\ref{sron}}
\and Manuel G\"udel\orcidlink{0000-0001-9818-0588}\inst{\ref{vienna},\ref{eth}} 
\and Thomas Henning\orcidlink{0000-0002-1493-300X}\inst{\ref{mpia}}
David Barrado\orcidlink{0000-0002-5971-9242}\inst{\ref{cab}}
\and Leen Decin\orcidlink{0000-0002-5342-8612}\inst{\ref{leuven}} 
\and John Pye\orcidlink{0000-0002-0932-4330}\inst{\ref{leicester}} 
\and Pascal Tremblin \orcidlink{}\inst{\ref{paris_maison}}
}
\institute{
 STAR Institute, Universit\'e de Li\`ege, All\'ee du Six Ao\^ut 19c, 4000 Li\`ege, Belgium\label{star_liege} 
\and Montefiore Institute, Universit\'e de Li\`ege, 10 All\'ee de la D\'ecouverte, 4000 Li\`ege, Belgium \label{mont_liege} 
\and  Max-Planck-Institut f\"ur Astronomie (MPIA), K\"onigstuhl 17, 69117 Heidelberg, Germany \label{mpia} 
\and ETH Z\"urich, Institute for Particle Physics and Astrophysics, Wolfgang-Pauli-Strasse 27, 8093 Z\"urich, Switzerland\label{eth} 
\and  Department of Astrophysics/IMAPP, Radboud University, PO Box 9010, 6500 GL Nijmegen, the Netherlands\label{radboud} 
\and  HFML - FELIX. Radboud University PO box 9010, 6500 GL Nijmegen, the Netherlands\label{hfml} 
\and  SRON Netherlands Institute for Space Research, Niels Bohrweg 4, 2333 CA Leiden, the Netherlands\label{sron} 
\and   Department of Astrophysics, University of Vienna, T\"urkenschanzstrasse 17, 1180 Vienna, Austria\label{vienna} 
\and Centro de Astrobiología (CAB), CSIC-INTA, ESAC Campus, Camino Bajo del Castillo s/n, 28692 Villanueva de la Cañada, Madrid, Spain \label{cab} 
\and Institute of Astronomy, KU Leuven, Celestijnenlaan 200D, 3001 Leuven, Belgium\label{leuven} 
\and  School of Physics \& Astronomy,  
Space Park Leicester, University of Leicester, 92 Corporation Road, Leicester, LE4 5SP, UK\label{leicester} 
\and Université Paris-Saclay, UVSQ, CNRS, CEA, Maison de la Simulation, 91191, Gif-sur-Yvette, France \label{paris_maison}
\and European Space Agency, Space Telescope Science Institute, Baltimore, MD, USA\label{esa} 
\and  Kapteyn Institute of Astronomy, University of Groningen, Landleven 12, 9747 AD Groningen, the Netherlands\label{kapteyn} 
\and  Institute for Astronomy, University of Edinburgh, Royal Observatory, Blackford Hill, Edinburgh EH9 3HJ\label{roe} 
\and Department of Astrophysics, American Museum of Natural History, New York, NY 10024, USA\label{museum} 
\and Jet Propulsion Laboratory, California Institute of Technology, 4800 Oak Grove Dr.,Pasadena, CA 91109, USA\label{jpl} 
\and Aix Marseille Univ, CNRS, CNES, LAM, Marseille, France\label{lam} 
\and Leiden Observatory, Leiden University, P.O. Box 9513, 2300 RA Leiden, the Netherlands\label{leiden} 
\and Cosmic Dawn Center (DAWN), DTU Space, Technical University of Denmark. Building 328, Elektrovej, 2800 Kgs. Lyngby, Denmark\label{dawn} 
\and Department of Astronomy, Oskar Klein Centre, Stockholm University, 106 91 Stockholm, Sweden\label{oskar} 
\and School of Cosmic Physics, Dublin Institute for Advanced Studies, 31 Fitzwilliam Place, Dublin, D02 XF86, Ireland\label{dublin} 
\and  Department of Physics and Astronomy, University College London, Gower Street, WC1E 6BT, UK\label{ucl} 
\and Jet Propulsion Laboratory, California Institute of Technology, Pasadena, California, United States\label{jpl}
\and Astronomical Institute 'Anton Pannekoek' Amsterdam
}
   \date{This work is under review at A\&A}
  \abstract
   {Cold brown dwarf atmospheres provide a good training ground for the analysis of atmospheres of temperate giant planets. WISEP~J173835.52+273258.9 (WISE 1738) is an isolated cold brown dwarf and a Y0 spectral standard with a temperature between 350–400~K, lying at the boundary of the T-Y transition. Although its atmosphere has been extensively studied in the near-infrared, its bulk physical parameters and atmospheric chemistry and dynamics are not well understood. }
   {Using a Mid-Infrared Instrument (MIRI) medium-resolution spectrum (5–18 $\mu$m), combined with near-infrared spectra (0.98–2.2 $\mu$m) from Hubble Space Telescope's (HST) Wide Field Camera 3 (WFC3) and Gemini Observatory's Near-Infrared Spectrograph (GNIRS), we aim to accurately characterize the atmospheric chemistry and bulk physical parameters of WISE~1738.}
   {We perform a combined atmospheric retrieval on the MIRI, GNIRS and WFC3 spectra using a machine learning algorithm called Neural Posterior Estimation (NPE) assuming a cloud-free model implemented using \texttt{petitRADTRANS}. We demonstrate how this combined retrieval approach ensures robust constraints on the abundances of major atmospheric species, the pressure-temperature ($P$-$T$) profile, bulk C/O and metalliclity [M/H], along with bulk physical properties such as effective temperature, radius, surface gravity, mass and luminosity. We estimate 1D and 2D marginal posterior distributions for the constrained parameters and evaluate our results using several qualitative and quantitative Bayesian diagnostics, including Local Classifier 2-Sample Test (L-C2ST), coverage and posterior predictive checks.
      }
   {The combined atmospheric retrieval confirms previous constraints on H$_2$O, CH$_4$, NH$_3$, and for the first time provides constraints on CO, CO$_2$ and $^{15}$NH$_3$. It also gives better constraints on the physical parameters and the $P$-$T$ profile, while also revealing potential biases in characterizing objects using data from limited wavelength ranges. The retrievals further suggest the presence of disequilibrium chemistry, as evidenced by the constrained abundances of CO and CO\(_2\), which are otherwise expected to be depleted and hence not visible beyond the near-infrared wavelengths under equilibrium conditions. We estimate the physical parameters of the object as follows: an effective temperature of 402\(^{+12}_{-9}\)~K, surface gravity (\(\log g\)) of 4.43\(^{+0.26}_{-0.34}\)~cm~s\(^{-2}\), mass of 13\(^{+11}_{-7}\)~\(M_{\text{Jup}}\), radius of 1.14\(^{+0.03}_{-0.03}\)~\(R_{\text{Jup}}\), and a bolometric luminosity of \(-6.52^{+0.05}_{-0.04}\)~\(\log L/L_\odot\). Based on these values, the evolutionary models suggest an age between 1 and 4~Gyr, which is consistent with a high rotation rate of 6 hr of the brown dwarf. We further obtain an upper bound on the \(^{15}\)NH\(_3\) abundance, enabling a 3$\sigma$ lower bound calculation of the \(^{14}\)N/\(^{15}\)N ratio = 275, unable to 
   interpret the formation pathway as core collapse. Additionally, we calculate a C/O ratio of 1.35\(^{+0.39}_{-0.31}\) and a metallicity of 0.34\(^{+0.12}_{-0.11}\) without considering any oxygen sequestration effects. 
    }
   {}
   \maketitle

%

\section{Introduction}

    Brown dwarfs serve as a bridge between planetary and stellar objects, making them essential for understanding planetary atmospheres and formation mechanisms.
    Their atmospheres resemble those of low-irradiation giant exoplanets characterized by low effective temperatures and complex molecular chemistry. However, what sets them apart from planets is their higher mass and surface gravity, with a few exceptions such as PSO~318 (\citealt{2013ApJ...777L..20L}), imparting them star-like characteristics. This provides interesting planet-brown dwarf analogs such as the recently characterized Eps Ind Ab \citep{2024Natur.633..789M} and WISE 0855 \citep{helena_2024}.
    
    Late T and Y dwarfs, the coldest and least luminous brown dwarfs, provide opportunities to study sub-stellar atmospheres akin to those of temperate to cold gas giants, without the complication of the need for high-contrast imaging in the presence of a bright host star. Although some works theoretically suspect clouds in these atmospheres \citep{morley2014water, 2022ApJ...927..184M}, a population-wide Y-dwarf retrieval analysis on HST data \citep{zalesky2019uniform} and more recently the longer wavelength range (JWST) data analysis of Y-dwarfs WISE~0855 \citep[300K,][]{helena_2024} and WISE~1828 \citep[380K,][]{barrado202315nh3} suggest a cloud-free atmosphere, dominated by strong H$_2$O and CH$_4$ absorption. 
    
    Among over 20 confirmed Y dwarfs \citep{2019ApJS..240...19K}, WISEP J173835.52+273258.9 (henceforth WISE 1738) was one of the first ultracool dwarfs identified using the Wide-field Infrared Survey Explorer \citep[WISE;][]{kirkpatrick2011first}. Its discovery was announced by \citet{cushing2011discovery}, with a spectrum obtained using the infrared channel of the WFC3 \citep{10.1117/12.789581} onboard the HST at a resolution of approximately $130$. With a calculated temperature of approximately $400$~K, it lies at the boundary between T and Y dwarfs. The steep flux drop in the blue wing of the $H$-band, associated with NH$_3$ absorption at 1.49~$\mu$m, led to its classification as a Y0 spectral standard \citep{cushing2011discovery}. Since then, it has been extensively studied in the near-infrared. Early atmospheric characterization was performed by comparing the data to self-consistent radiative-convective-thermochemical equilibrium models \citep[e.g.,][]{1996ApJ...465L.123A, 1996Sci...272.1919M, 1996A&A...308L..29T, 2001RvMP...73..719B}. An early spectral fit by \citet{schneider2015hubble} using cloud-free and chloride, sulfide and water cloud models from \citet{saumon2012new,Morley_2012,morley2014water} on HST/WFC3  data, did not achieve satisfactory agreement between models and observations. The discrepancies were attributed to the assumption of equilibrium chemistry, which can significantly affect estimates of effective temperature, surface gravity, and cloud properties 
    
    Subsequent higher-resolution ($\lambda/\Delta\lambda \approx  2800$) near-infrared spectra from instrument GNIRS \citep{10.1117/12.671817} at the Gemini Observatory highlighted the critical role of disequilibrium chemistry \citep{leggett2015near, leggett2016near, leggett2017type} based on model comparisons with \citet{saumon2012new, 2015ApJ...804L..17T, Morley_2012, morley2014water}. Additionally, light curve variability (approximately $3\%$) observed in the Spitzer 4.5~$\mu$m band, corresponding to a rotation period of $6.0 \pm 0.1$~hr, was attributed to patchy KCl and Na$_2$S clouds \citep{leggett2016observed}. Cloud-free models generally still provided better fits, especially for objects with T eff between 400 K and 450 K, although the Y-band still showed poor fits, likely due to variations in metallicity, surface gravity, and cloud properties.

    While grid models describe planetary atmospheres using a few fundamental parameters, such as effective temperature and surface gravity, they generally rely on assumptions to predict thermal structures and to determine molecular abundances. However, these models fail to capture dynamical processes hinting at disequilibrium chemistry. In contrast, atmospheric retrieval methods, first applied to exoplanet studies by \citet{2009ApJ...707...24M}, invert observed spectra to directly infer temperature structures and molecular abundances with minimal prior assumptions, thus describing the atmosphere with more parameters. While this flexibility can sometimes lead to nonphysical results, retrievals provide unique insights into complex atmospheric processes that grid models cannot fully capture. In short, there is a fundamental trade-off between the physical self-consistency of grid models and the empirical flexibility of free retrievals. 
    
    
    \citet{zalesky2019uniform} conducted free retrievals on the HST spectrum of WISE 1738 incorporating disequilibrium chemistry. The retrieved $P$-$T$ profile was consistent with radiative-convective equilibrium and suggested a cloud-free atmosphere, even though the $P$-$T$ profile intersected Na$_2$S and KCl condensation curves within the photosphere and showed water condensates at higher altitudes. However, the retrieved parameters were inconsistent with evolutionary models, indicating unusually high surface gravity and mass.
    
    In this study, we investigate the atmosphere of the Y0 spectral standard WISE 1738 using, for the first time, a mid-infrared spectrum obtained with JWST's MIRI instrument, alongside near-infrared data from HST/WFC3 and Gemini/GNIRS. The mid-infrared region probes higher altitudes in the atmosphere compared to the previously studied near-infrared region, offering a new perspective on the atmosphere. This allows us to place improved constraints on bulk physical properties such as surface gravity, radius, mass, and luminosity, while also revealing the potential prevalence of disequilibrium chemistry due to a more robust understanding of its chemical composition. The paper is organized as follows: Section~\ref{sec:data_processing} describes the data processing for the three datasets. In Section~\ref{sec:simulator} we describe the model used for the retrieval. In Section~\ref{sec:retrieval}, we perform a combined retrieval on the datasets (HST/WFC3, Gemini/GNIRS, and MIRI/JWST) using a simulation-based inference algorithm. Section~\ref{sec:retrieval_results} presents the retrieval results, which are further analyzed and discussed in Section~\ref{sec:discussion}. Finally, Section~\ref{sec:conclusion} provides concluding remarks.

\section{Data processing}
\label{sec:data_processing}

    \subsection{JWST/MIRI reduction}
        The MIRI instrument onboard the JWST \citep{2015PASP..127..595W, 2023PASP..135d8003W} includes the Medium Resolution Spectrometer \citep[MRS,][]{2023PASP..135d8003W}, which provides medium-resolution spectroscopy over the mid-infrared wavelength range of approximately 5 to 28 $\mu$m \citep{2023A&A...675A.111A}. The presented MRS data are part of the Guaranteed Time Observation (GTO) program ``MIRI Spectroscopic Observations of Brown Dwarfs'' under the observation ID: 1278 led by Pierre-Olivier Lagage. The MIRI/MRS data of WISE 1738 were obtained on July 18th 2023 at 17:44:58 UTC. The observation ran in the FASTR1 read out pattern and the two point dither pattern with four exposures, one integration per exposure and 110 groups per integration.
        The time between each frame was 2.78~s, as well as the time between groups, resulting in a total exposure time of 610.5~s. During the mid-time of the exposure, the telescope pointed at RA 17$^{\mathrm{h}}$\,38$^{\mathrm{m}}$\,24$^{\mathrm{s}}$ and DEC +27$^\circ$\,30$'$\,0$''$.

        
        The data were downloaded from the Mikulski Archive for Space Telescopes (MAST, DOI: \href{https://mast.stsci.edu/portal/Mashup/Clients/Mast/Portal.html?searchQuery=%7B%22service%22:%22DOIOBS%22,%22inputText%22:%2210.17909/vzcg-p593%22%7D}{10.17909/vzcg-p593}) and processed through the pipeline. The pipeline \citep{2023zndo...7577320B} consists of multiple stages, where the first stage includes the ramp fitting to calibrate the raw data to flux units. The second stage assigns a world coordinate system, applies a flat field, straylight, fringe and photometric correction. After this step, the background is removed by subtracting the two dithers from each other (for details, see \citealt{barrado202315nh3}). The last stage converts the detector data set to a 3D cube using the 'drizzle' weighting algorithm after assigning a coordinate system to it and running an outlier detection to flag remaining bad pixels. In the end, the spectrum is extracted by centring an aperture at the source with a radius of one full width at half maximum (FWHM) of the  point spread function (PSF) of the object. An additional one dimensional fringe correction in the extraction function is performed to reduce additional fringing effects. The data reduction was done using the JWST pipeline version 1.12.5, CRDS version 11.17.10 and context file version jwst$\_$1149.pmap.
        
\begin{figure*}[!t]
    \centering
    \includegraphics[width=\textwidth]{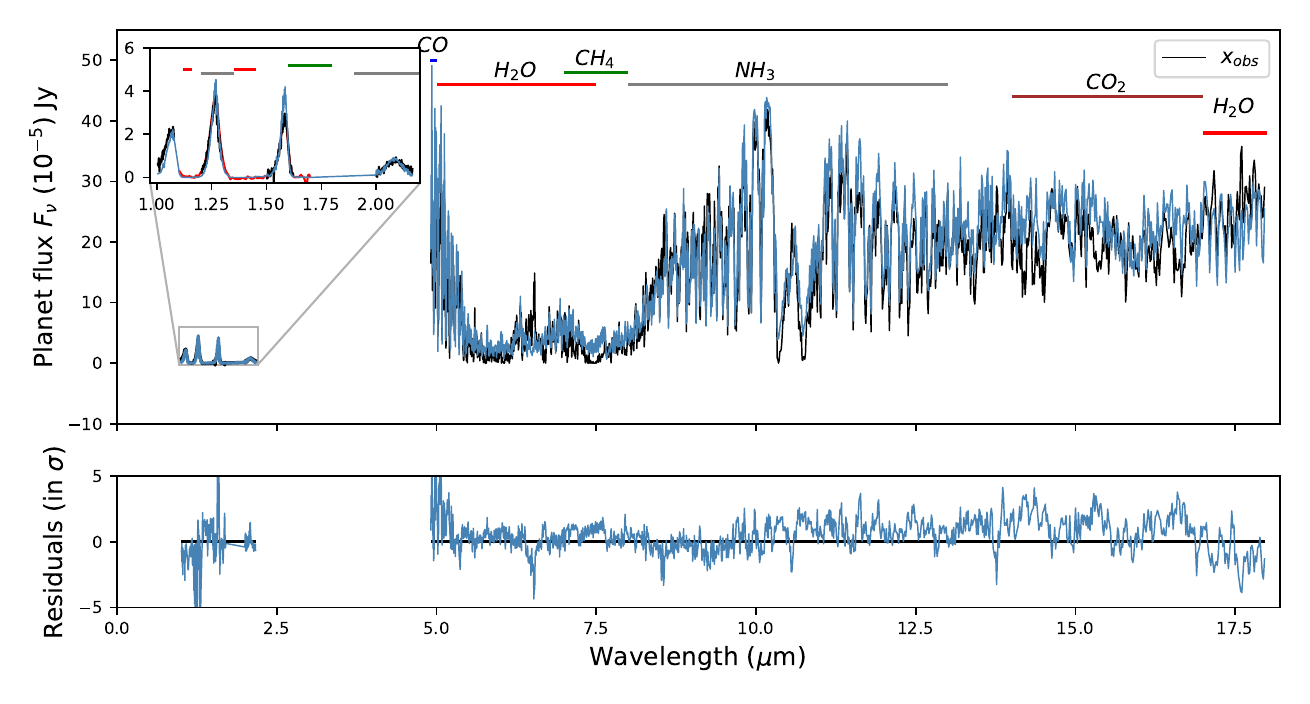}
    \caption{\textit{Top.} WFC3 ($J$ and $H$ bands, red), GNIRS ($Y$, $J$, $H$ and $K$ bands, black), and MIRI (black) observations $x_\text{obs}$, overlaid with the simulated noiseless spectrum $f(\theta)$ associated with the most probable parameters from the posterior. {\textit{Bottom.} Residuals of the sample normalized by the inflated standard deviation of the noise distribution for each spectral channel.
    }}
    \label{fig:cloudfree_mostprob}
    \end{figure*}

        The output MIRI spectrum was obtained from channels 1A to 3C, ranging between 4.9 $\mu$m and 17.9 $\mu$m. Each channel was rebinned to the \texttt{petitRADTRANS} wavelength spacing of $\lambda/\Delta\lambda = 1000$. 
        Accounting for the overlaps at the edges of each channel by averaging over the repeated wavelengths, and stitching them all together, resulted in the final vector used for retrievals. The spectrum is shown in Fig.~\ref{fig:cloudfree_mostprob} in black. Visual inspection already suggests the unambiguous presence of molecules such as CO, H$_2$O, CH$_4$, NH$_3$ and CO$_2$ as indicated in the figure. 
        
    \subsection{HST/WFC3 spectrum}

        The near-infrared spectrum for WISE 1738 was obtained from the data observed on the infrared channel of the WFC3 \citep{10.1117/12.789581} on-board the HST, as a part of its Cycle 18 program (GO-12330, PI: J. D. Kirkpatrick) in 2011. Further details on how it was acquired can be found from the discovery paper by \citet{cushing2011discovery}. The HST spectrum covers a wavelength range  of 1.07–1.70 $\mu$m at a resolving power of approximately $130$. 
        The spectrum was obtained from Figure 8 of \citet{schneider2015hubble}. 

    \subsection{Gemini/GNIRS spectrum}
    
        A higher-resolution near infrared spectrum for WISE 1738 was also obtained from the data obtained by the Gemini North telescope using the GNIRS, at a resolving power of approximately $2800$, via the program GN-2014A-Q-64 \citep{elias2006performance}. Further information on this can be found in the work by \citet{leggett2016near}. The spectrum covers a wavelength range of 0.993–1.087 $\mu$m, 1.191–1.305 $\mu$m, 1.589–1.631 $\mu$m and 1.985–2.175 $\mu$m, i.e., spanning the \textit{Y, J, H} and \textit{K} bands.
        For the retrievals in this work, this spectrum is rebinned down to a resolution $\lambda/\Delta\lambda = 1000$ to match the default resolution of \texttt{petitRADTRANS}.  
        
\section{Atmospheric radiative transfer model}
    \label{sec:simulator}

    The atmospheric forward model $f$ used in this study is implemented using \texttt{petitRADTRANS} (version 2.6.3) along with a noise model to account for the measurement noise. \texttt{petitRADTRANS} \citep{molliere2019petitradtrans} is a one-dimensional radiative transfer model that is used to calculate the emission and transmission spectra for exoplanets with cloudy and cloud-free atmospheres. This model simulates a single atmospheric column consisting of multiple distinct pressure layers, with temperatures determined by a freely parameterized thermal profile. The layers include various opacity sources dispersed throughout the column, and are incorporated into the radiative transfer equations. Given that objects at this temperature are sufficiently cold for silicate clouds to have dispersed and have been well characterized without the need for water or ammonia clouds \citep{2024arXiv241010933K, barrado202315nh3, zalesky2019uniform}, we adopt this setup to model a cloud-free atmosphere.
    
    The cloud-free model is parameterized using 26 parameters, denoted as $\theta$. The $P$-$T$ profile is calculated on a pressure grid containing various levels between $10^{-6}$ and 1000 bar. Within this grid, 10 nodes are equidistantly defined in log-space. The temperature at the bottom-most node is set as a free parameter $T_{\rm bottom}$, such that it can take any value  uniformly distributed between 100 to 9000K. The temperature at each upper node is calculated as a parameterized fraction uniformly distributed between 0.2-1.0 of the temperature at the node immediately below it. These 9 node fractions are defined as $T_{\rm{nodes[i-ix]}}$. Once the temperatures at all the nodes are calculated, the entire profile is constructed by quadratically interpolating between them. 
    
    The primary absorber species typical in Y-dwarf atmospheres with an effective temperature (\( T_{\rm eff} \)) of approximately 400 K—such as CH\(_4\), H\(_2\)O, H\(_2\)S, CO\(_2\), CO, and NH\(_3\)—are considered \citep[see Figure 5 in][]{leggett2015near}, along with the isotopologue \({}^{15}\)NH\(_3\) \citep{barrado202315nh3}. Our model additionally incorporates HCN and PH$_3$ \citep{Visscher_2006, Zahnle_2014} and TiO and VO, the latter two of which are notably present in higher temperature brown dwarfs, such as late-type M dwarfs \citep{1999ApJ...519..802K}. Continuum species such as H$_2$-H$_2$ and He-H$_2$ collision-induced absorption bands are also considered. The logarithm (of base 10) of the abundance of each opacity species, expressed as mass fractions, is treated as a free parameter. Brown dwarf atmospheres are expected to exhibit significant turbulence, affecting species such as CH\(_4\), CO, CO\(_2\), N\(_2\), and NH\(_3\) \citep[e.g.,][]{1997ApJ...489L..87N, 2000ApJ...541..374S, 2004AJ....127.3516G, 2007ApJ...655.1079L, 2011ApJ...738...72V, 2014ApJ...797...41Z}. Further \citet{2022ApJ...938..107M} suggest (log) mixing values between 6-7 for brown dwarfs between 400-500~K. Consequently, we assume the considered abundances remain constant throughout the pressure column, with values (in log10 units) uniformly distributed between \(-10\) and \(0\). We note that recent studies such as \citet{2023ApJ...947....6R} have demonstrated that this approach may oversimplify atmospheric complexities, potentially leading to inaccurate retrievals of gravity, metallicity, and C/O ratios, irrespective of the parameterization of the $P$-$T$ profile. This could be incorporated into future retrievals that utilize broad wavelength ranges enabled by combined approaches. 
    
    The spectra are calculated with the radiative transfer routines implemented in  \texttt{petitRADTRANS}. The planet mass $M_p$ and radius $R_p$ are also considered to be free parameters and are used to calculate the emission flux. 
     Measurement noise $\epsilon$ is added to the generated flux $f(\theta)$ such that the simulator output is given as $x = f(\theta) + \epsilon$, where $\epsilon$ is drawn from a Gaussian noise distribution $\mathcal{N}(0, \sigma_N^{2})$ defined uniquely for each instrument. Here, $\epsilon \in \mathbb{R}^{L}$ represents a vector containing random noise instances in each wavelength bin, where $L$ is the combined spectral length of the WFC3+GNIRS+MIRI observations. To account for the random instrument effects and the missing forward model physics, an additional scaling factor $b$ is added to inflate the standard deviation on the noise measured for each instrument such that the total error $s$ is given by $s^{2} = \sigma_N^{2} + 10^{b}$ \citep{Line_2015}.  The free parameters $b_{w}$, $b_{g}$ and $b_{m}$ pertaining to instruments WFC3, GNIRS and MIRI instruments respectively, are set to take values uniformly distributed in the range of [-17, -11), [-17, -11) and [-15, -7). These parameter ranges scale the maximum standard deviations of their respective instruments by factors of 1.35, 1.1 and 33 times respectively. A lower prior range chosen for the WFC3 and GNIRS than for the MIRI $b$ factors is motivated by relatively robust and well-understood (preliminary) error bars on these instruments. 
     
     
     The prior distribution in the case of a cloud-free model is a 26-dimensional multivariate uniform distribution $p(\theta)$ with physically motivated ranges for each parameter listed in Table~\ref{tab:prior}.


\section{Atmospheric retrieval}
\label{sec:retrieval}
    \begin{table}[t]
    \caption{Prior distribution for the 26 model parameters.}\label{tab:prior}
    \centering
    \small
    \begin{tabular}{cc|cc}
        \hline \hline
        Parameter & Prior & Parameter & Prior\\
        \hline
        $R_{P}$ & $\mathcal{U}[0.5, 3)$ & H$_2$O & $\mathcal{U}[-10, 0)$ \\
        $M_{P}$ & $\mathcal{U}[1, 50)$ &  CO$_2$ & $\mathcal{U}[-10, 0)$ \\ 
        $T_{\rm bottom}$ ~\tablefootmark{a} & $\mathcal{U}[100, 9000)$ & CO & $\mathcal{U}[-10, 0)$ \\ 
        $T_{\rm nodes[i-ix]}$~\tablefootmark{b} & $\mathcal{U}[0.2, 1)$ & CH$_4$ & $\mathcal{U}[-10, 0)$ \\
        $b_{w} ~\tablefootmark{c}$  & $\mathcal{U}[-17, -11)$ & NH$_3$ & $\mathcal{U}[-10, 0)$ \\ 
         $b_{g}$ ~\tablefootmark{c} & $\mathcal{U}[-17, -11)$ & PH$_3$ & $\mathcal{U}[-10, 0)$ \\ 
        $b_{m} ~\tablefootmark{c} $&  $\mathcal{U}[-15, -7)$ & H$_2$ & $\mathcal{U}[-10, 0)$ \\ 
         &  & $^{15}$NH$_3$ & $\mathcal{U}[-10, 0)$  \\
         & & HCN & $\mathcal{U}[-10, 0)$ \\
         & &  TiO & $\mathcal{U}[-10, 0)$ \\
         & & VO & $\mathcal{U}[-10, 0)$ \\ 
        \hline
    \end{tabular}
    \tablefoot{
        All the abundances are mass fractions in $\log_{10}$ units. \\
        \tablefoottext{a}{$T_{\rm bottom}$ is the temperature at the bottom-most node in the pressure grid. }
        \tablefoottext{b}{$T_{\rm nodes[i-ix]}$ are the subsequent fractions of the previous node temperatures. }
        \tablefoottext{c}{The $b$ factor for the instruments are additive noise factors, in log value, by which the square of the measured error bars are exaggerated in each bin of the spectrum. This embodies the uncertainty in the estimated error of each instrument or model inaccuracies.}
    }
    \end{table}

        We perform the atmospheric retrieval using a simulation-based inference algorithm called neural posterior estimation \citep[NPE,][]{2023A&A...672A.147V}. In NPE, a neural network based model called a conditional normalizing flow is trained to estimate the Bayesian posterior distribution $p(\theta | x)$ over the model parameters $\theta$ given the observation $x$ 
     (see \citet{2023A&A...672A.147V} for further details on the setup). The training phase includes sampling from the prior distribution $p(\theta)$ over the model parameters, and passing it through an atmospheric simulator $p(x|\theta)$ (forward model+noise model) in order to generate noisy spectral simulations. These simulations are compressed in an embedding network, to avoid overfitting by memorization, and used to train the normalizing flow to generate a conditional probability distribution  $p_\phi(\theta|x)$. During inference, the trained normalizing flow conditioned on an observation $x_{\rm obs}$, is used to obtain an approximation of the posterior distribution. 



    \begin{figure*}[!t]
    \centering
    \includegraphics[width=\textwidth]{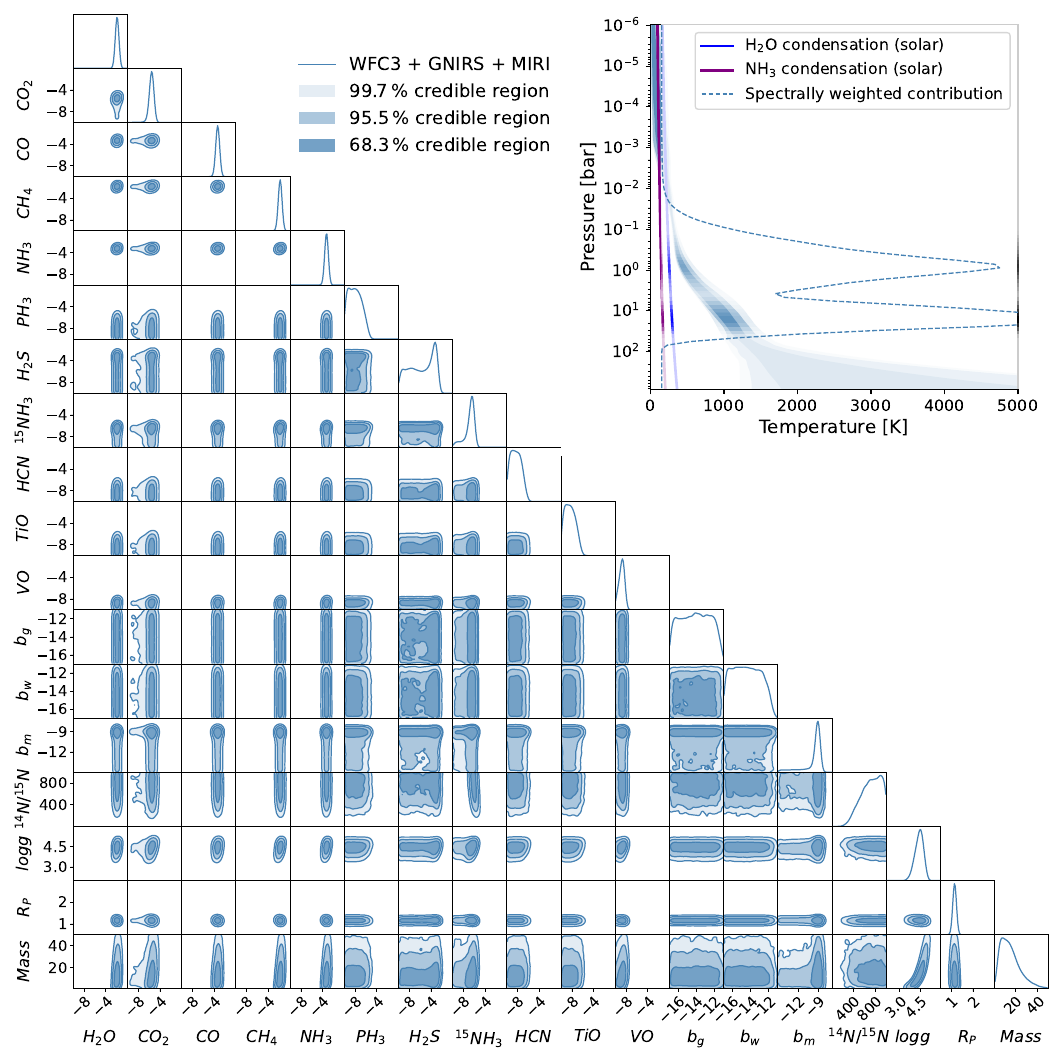}
    \caption{\textit{Left}. Cloud-free retrieval using neural posterior estimation on the WISE 1738 spectra. The corner plot shows the full 1D and 2D marginal posterior distributions obtained for the WISE 1738 observations $x_\text{obs}$, WFC3+GNIRS+MIRI spectra.} {\textit{Right.} The top right figure illustrates the posterior distribution of the $P$-$T$ profiles, that has the emission contribution function overlaid on top of it in shades of white highlighting visibility. The equilibrium state water ice and ammonia condensation curves are plotted along the profile in blue and purple respectively for solar metallicity [M/H] and C/O ratio. }
    \label{fig:cloudfree_corner_full}
    \end{figure*}

     The training set consists of approximately 3.7 million pairs of parameters and spectra $(\theta, f(\theta))$, providing around 1.8 value points per dimension if one were to use a regular 26-dimensional grid. The sample pairs in this training set are split into 90\%, 9\% and 1\% for training, validation and testing respectively. The training set includes combined simulations of the near-infrared and the mid-infrared wavelengths.
      For the training set, atmospheric models between the wavelength ranges of $0.98-2.2$~$\mu$m in the near-infrared and $4.9-18$~$\mu$m in the mid-infrared are simulated to match the default wavelength spacing of $\lambda/\Delta\lambda = 1000$ in \texttt{petitRADTRANS}. The WFC3 and GNIRS spectra have overlapping coverage in the near-infrared region, although they have different wavelength spacings. To match the WFC3 wavelengths, the spectra in the NIR range are convolved to the WFC3 spectral resolution and rebinned to a spacing of 130. 
      To generate the GNIRS component of the training set, the NIR spectra at the \texttt{petitRADTRANS} simulated default wavelength spacing of $\lambda/\Delta\lambda = 1000$ is retained, but masked to match the rebinned GNIRS observations. Similarly, the mid-infrared spectra are generated to match the MIRI observations. All three component spectra are combined to generate the training set. The simulations took around 3200 CPU hours.
      
      
    During training, random noise realizations are added on-the-fly to the simulated spectra in the training set, to obtain simulated observations $x = f(\theta) + \epsilon$. These joint pairs of noisy simulations and their corresponding model parameters are input into the normalizing flow. 
    In each forward pass, 
    the posterior log-density $\log p_\phi(\theta | x)$ is computed, where $\phi$ is the set of weights of the neural flow network, and the loss function $\iint -\log p_\phi(\theta | x) \, p(\theta, x) \, d\theta \, dx$ on $\phi$ is minimized over the whole training set ($\theta, x $) \citep{papamakarios2021normalizing}. Hyper-parameter tuning is performed by conducting 128 parallel runs, each with a configuration randomly chosen from a uniform grid (see Appendix~\ref{sec:NFarch}) over all the hyper-parameters. The run leading to lower validation loss and/or more stable training across all atmospheric models is preferred. Each hyper-parameter run takes around 24 hours. 
    The time taken to train the posterior estimator $p_{\phi}(\theta|x)$ is approximately 5.8 hours. Sampling 20,469 times from the normalizing flow (i.e., the posterior estimate) took approximately 1 minute on average. Therefore the total time required to perform a single retrieval, including the overhead cost of dataset set generation, hyper-parameter tuning, and training was approximately 30 hours. The technical details of the architecture of the normalizing flow, hyper-parameter tuning and the training procedure are provided in Appendix ~\ref{sec:exp-diagnostics}. 


\section{Retrieval results}
    \label{sec:retrieval_results}

        \begin{table*}[!t]
    \caption{\label{table:atmospheric_abundances}Retrieved atmospheric (log) abundances as volume mixing ratios.}
    \centering
    \begin{tabular}{lccccccc}
    \hline\hline
    Retrieval & Instrument & H$_{2}$O & CH$_{4}$ & CO & CO$_2$ & NH$_3$ \\
    \hline
    Z19-free$^{a}$ & WFC3 & \(-2.87^{+0.08}_{-0.08} \) & \(-2.75^{+0.12}_{-0.10} \) & \(-3.3 \) & \(-4.1 \) & \(-4.21^{+0.10}_{-0.09} \) \\
        
    Z19-constr.$^{b}$ & WFC3 & \(-2.97^{+0.09}_{-0.12} \) & \(-2.89^{+0.12}_{-0.13} \) & \(-3.79 \) & \(-3.83 \) & \(-4.34^{+0.12}_{-0.13} \) \\
        
    This work$^{c}$ & WFC3 + GNIRS + MIRI & \(-2.86^{+0.11}_{-0.11} \) & \(-2.72^{+0.14}_{-0.15} \) & \(-4.49^{+0.18}_{-0.18} \) & \(-6.87^{+0.25}_{-0.31} \) & \(-4.2^{+0.12}_{-0.12} \) & \\
    \hline
    \end{tabular}
    \tablefoot{Log abundances are expressed units of volume mixing ratios. \\
    \tablefoottext{a}{Called the ``free'' model with 31 parameters. This  incorporates a 80 $M_{\text{Jup}}$ mass prior upper limit. The 3$\sigma$ upper limits are from Table 4 of \citet{zalesky2019uniform}.}
    \tablefoottext{b}{Called the ``constrained'' model, also with 31 parameters. Here the previously used upper mass limit is removed and restraints are applied on the priors of radius and $\log g$, as $0.7 < R/R_{\text{Jup}} < 2.0$ and $3.5 < \log(g) < 5.5$ respectively. The 3$\sigma$ upper limits are computed from the marginal posteriors obtained by \citet{zalesky2019uniform}.}
    \tablefoottext{c}{3$\sigma$ upper limits computed from the marginal posteriors obtained in our work. The volume mixing ratios in this work were calculated by converting the mass fractions from the cloud-free retrieval.}
    }
    \end{table*}

\begin{table*}[!t]
    \caption{\label{table:summary_data_models}Summary of previous model fits/retrievals attempting to characterize WISE 1738.}
    \centering
    \begin{tabular}{lccccccccc}
    \hline\hline
    Study & $T_{\rm eff}$ (K) & $\log g$ (cm/s$^2$) & Mass ($M_{\rm Jup}$) & Radius ($R_{\rm Jup}$) & Age (Gyr) & $\log K_{\text{zz}}$ & Distance (pc) & C/O & [M/H] \\
    \hline
    C11 & 350--400 & 4.75--5.0 & 20 & 0.86--0.94 & -- & 4 & 3.4--7.3$^{d}$ & -- & solar \\
    S15 & 400 & 4.0--4.5 & 5--14$^{a}$ & 0.47$^{b}$ & 0.6--3 & -- & -- & -- & solar \\
    L16b & $425 \pm 25$ & $4.0 \pm 0.25$ & 3--9$^{a}$ & -- & 0.15--1 & 6 & $7.8 \pm 0.6^{e}$ & -- & 0, +0.2 \\
    L17 & 410--440 & 4.0--4.5 & 5--14$^{a}$ & 1.0--1.2$^{a}$ & 0.3--3 & 6 & $10.5^{f}$ & -- & $-0.05\pm 0.25$ \\
    Z19-free & $371^{+27}_{-29}$ & $5.43^{+0.13}_{-0.17}$ & $59^{+15}_{-22}$ & $0.71^{+0.05}_{-0.05}$ & >10 & -- & $7.34 \pm 0.22^{g}$ & $1.32^{+0.1 h}_{-0.09}$ & $0.35^{+0.10 h}_{-0.09}$ \\
    Z19-constr. & $371^{+33}_{-30}$ & $5.20^{+0.20}_{-0.29}$ & $34^{+20}_{-17}$ & $0.73^{+0.04}_{-0.03}$ & >10 & -- & $7.34 \pm 0.22^{g}$ & $1.2^{+0.09 h}_{-0.03}$ & $0.23^{+0.11 h}_{-0.13}$ \\
    This work & $402^{+12}_{-9}$ & $4.43^{+0.26}_{-0.34}$ & $13^{+11}_{-7}$ & $1.14^{+0.03}_{-0.03}$ & 1--4$^{c}$ & -- & $7.34 \pm 0.22^{g}$ & $1.35^{+0.39}_{-0.31}$ & $0.34^{+0.12}_{-0.11}$ \\
    \hline
    \end{tabular}
    \tablefoot{
    The studies by \citet{cushing2011discovery} (C11), \citet{schneider2015hubble} (S15), and \citet{zalesky2019uniform} (Z19) utilize data from WFC3 \citep{cushing2011discovery} to perform their analysis, whereas \citet{leggett2016near} (L16b) and \citet{leggett2017type} (L17) rely on data from GNIRS \citep{leggett2016observed}. \\
    \tablefoottext{a}{Estimated from the evolutionary model of \citet{saumon2008evolution}.}
    \tablefoottext{b}{Computed by multiplying the retrieved ratio 6.445$\times$ $10^{-2}$ R$_{\text{Jup}}$/pc  by the distance measure from \citet{2018ApJ...867..109M}.}
    \tablefoottext{c}{The age is estimated using the evolutionary model from \cite{ 2021ApJ...920...85M}.}
    \tablefoottext{d}{Distance measure from \citet{cushing2011discovery}.}
    \tablefoottext{e}{Distance measure from \citet{kirkpatrick2011first}.} 
    \tablefoottext{f}{Distance measure from \citet{Beichman_2014}.} 
    \tablefoottext{g}{Distance measure from \citet{2018ApJ...867..109M}.} 
    \tablefoottext{h}{Computed from the posteriors of \citep{zalesky2019uniform} without sequestration.}
    
    }
    \end{table*}

     The results of the NPE retrievals on the combined  WFC3+GNIRS+MIRI spectral observations $x_\text{obs}$ of WISE~1738 are presented in the form of 1D and 2D marginal posterior distributions in Fig.~\ref{fig:cloudfree_corner_full}. These are approximated by extensively sampling from the estimated joint posterior distribution by means of forward passes through the trained normalizing flow. These samples are then used to construct the 68.3\%, 95.5\% and 99.7\% credible regions of the $P$-$T$ profile. Along with the constrained parameters, we also plot the derived (i.e, not retrieved) posterior distributions of the $^{14}$NH$_{3}$$\slash$$^{15}$NH$_{3}$ ratio and $\log g$. The figure also includes the posterior distribution of the $P$-$T$ profile, shown in the inset. The emission contribution function is overlaid on top of the $P$-$T$ profile in dashed lines, brightly highlighting the regions of the atmosphere that are probed by the observations. It also includes the equilibrium state water ice and ammonia, metal chloride and sulfide condensation curves \citep{2006asup.book....1L} in blue and purple, respectively. It can be seen that the water condensation curve intersects with the $P$-$T$ profile above the upper limit of the probed photosphere.

    We obtain clear constraints on the abundances of H$_2$O, CO$_2$, CO, CH$_4$, and $^{15}$NH$_3$. All the constrained abundance values are documented in Table~ \ref{table:atmospheric_abundances}. We identify upper bounds on PH$_3$, H$_2$S, HCN, TiO and VO implying a non-robust detection \citep{Line_2015, Line_2017}. We also observe that the $^{14}{\rm NH}_3/ ^{15}{\rm NH}_3$ ratio is not well constrained, and gives a lower bound of 200. All the retrieved and computed physical parameters are documented in Table~ \ref{table:summary_data_models}. The most probable sample from the posterior, and its normalized residuals which is normalized to the inflated standard deviation are displayed in Fig.~\ref{fig:cloudfree_mostprob}.

    The noise scaling factors (or $b$ factors) are barely constrained for the WFC3 and GNIRS spectra. This implies that the retrieval favors no specific values of noise scaling for the two spectra, within the chosen prior range, and suggests that any random effects are well accounted for within the uncertainties of other parameters without requiring additional noise adjustments. The drop in marginal values from $-$12 to $-$11 is a real effect that persists even when using a broader prior range. Consequently, we conclude that the $b$ factors for WFC3 and GNIRS have an upper bound at $-$12, reinforcing the robustness of these measurements. However, for MIRI, the error estimates are scaled with a $b$ factor around $-$9. This scaling factor implies that the uncertainty is approximately 4 times higher in the largest error bar compared to the measured value. 

     A detailed evaluation of the estimated posterior, including both qualitative and quantitative assessments such as coverage, posterior predictive distribution, and the L-C2ST test, is presented in Appendix~\ref{sec:exp-diagnostics}.
     
\section{Discussion}
\label{sec:discussion}
    \subsection{Combined retrieval vs near-infrared retrievals}
    \label{sec:combined_retrievals}

    \begin{figure*}[h!]
    \centering
    \includegraphics[width=\textwidth]{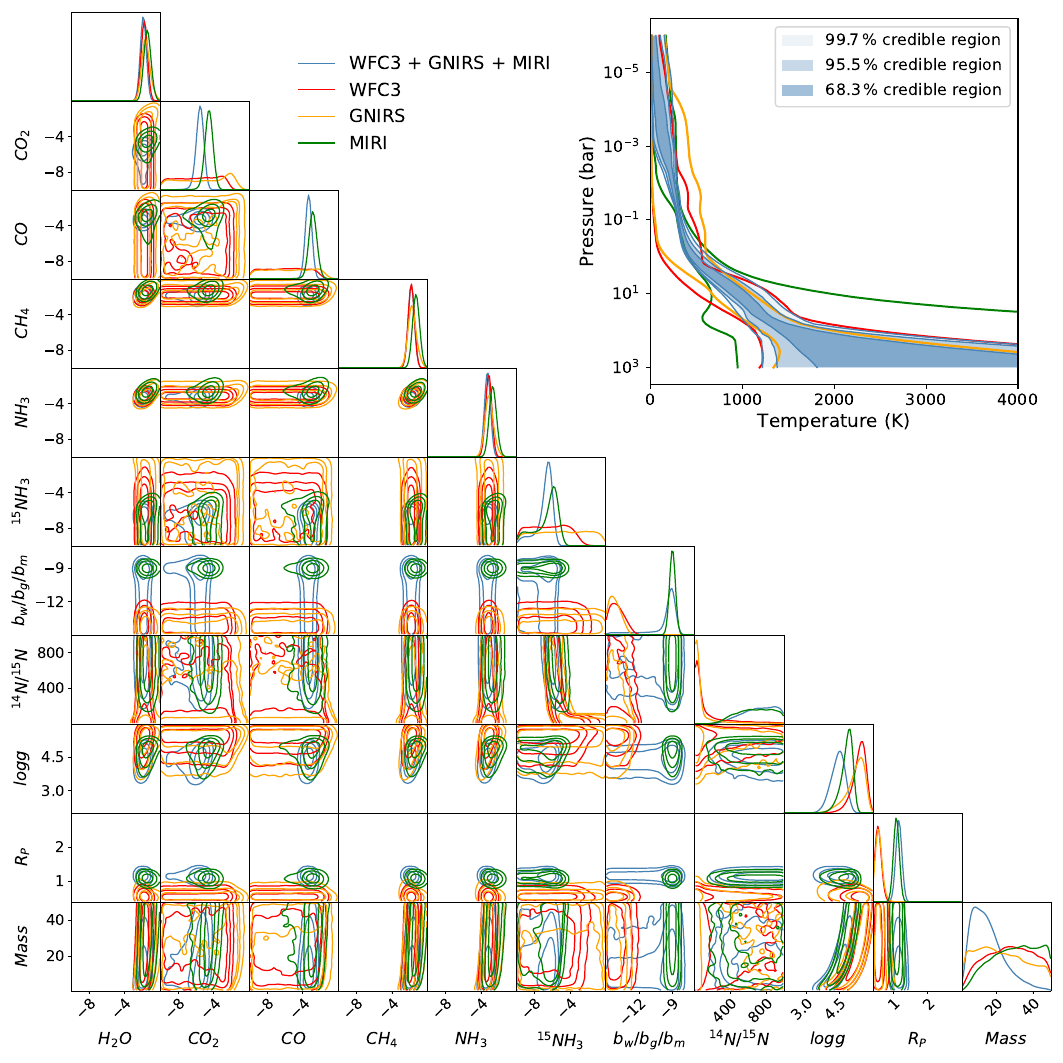}
    \caption{Comparing individual spectral retrievals of WISE 1738 across different wavelength regions with the combined retrievals. The corner plot shows 1D and 2D marginal posterior distributions obtained for the WFC3, GNIRS and MIRI spectrum along with the combined WFC3+GNIRS+MIRI spectra. The top right figure illustrates the posterior distribution of the $P-T$ profile of the combined retrieval, while also highlighting the 99.7\% credible intervals of the three independent retrievals. } 
    \label{fig:cloudfree_MIRI_HST_Gem_HST+Gem+MIRI}
    \end{figure*}

    Previous studies, such as those by \citet{cushing2011discovery, schneider2015hubble, zalesky2019uniform}, used WFC3 data to investigate the atmosphere of WISE 1738  while \citet{leggett2016observed, leggett2017type} used the GNIRS data for their analysis. Here, we combine these near-infrared datasets with mid-infrared observations for the first time. The combined retrieval of the WFC3, GNIRS, and MIRI spectra provides a more comprehensive view of the atmosphere by effectively expanding the range of probed pressure levels. This fills gaps even within the near infrared wavelength range by considering data at different resolutions/observational conditions. The advantage of combining datasets is demonstrated through a comparative study, where we perform single retrievals on WFC3, GNIRS and MIRI data separately, as well as a combined retrieval on WFC3, GNIRS, and MIRI data. The results of this comparison are shown in Fig.~\ref{fig:cloudfree_MIRI_HST_Gem_HST+Gem+MIRI}. 

    The comparison reveals that the abundances of CO, CO$_2$ and $^{15}$NH$_3$ are more tightly constrained in the combined retrievals than in those based solely on the near-infrared range. This is discussed in more detail in the following subsections. Additionally, although not significant, we see a similarly slight improvement in the confidence of constraints on the remaining abundances, suggesting that these features are strong in all datasets and hence easier to constrain. Similarly, the constraints on the $P$-$T$ profile also improve, with the near-infrared region providing tighter constraints on the deeper atmospheric layers, while the mid-infrared region contributes to tighter constraints in the upper layers. The most significant improvement however, is in the constraints on mass, radius and surface gravity. The free retrievals conducted exclusively on the near-infrared spectrum, both in our study and in the work by \citet{zalesky2019uniform}, result in higher estimates for mass (40 M$_J$) and surface gravity (5.5 cms$^{-2}$), along with a lower estimate for radius (0.7 R$_J$). In contrast, the combined retrievals result in lower masses and gravities, and larger radii that align better with the predictions from the evolutionary models (discussed more in Sect.~\ref{sec:evolution}). To explain this difference, we produce consistency plots across the entire wavelength range (see Figs.~\ref{fig:consistency_onlyMIRI},~\ref{fig:consistency_onlyGemini} and~\ref{fig:consistency_onlyHST} in Appendix~\ref{sec:exp-diagnostics}), which reveal that in each case, while retrievals remain consistent within the originally retrieved spectral region, they fail to predict consistent spectra in the extended (not retrieved) regions. This is in contrast with the consistency plot obtained for the combined retrieval in Fig.~\ref{fig:cloudfree_consistency} which exhibits consistency across the extended wavelength range. The enhanced consistency obtained in the latter case demonstrates how biased our characterizations can become when data coverage is limited. It also suggests that when focusing on narrow wavelength regions, multiple competing hypotheses can fit the spectral shape. 
    Therefore, the MIRI data provide critical additional information, enabling improved constraints on these physical parameters. These findings highlight the necessity of incorporating broader datasets to achieve accurate characterizations of brown dwarfs. 
    
    Interestingly in the combined retrievals, while the normalized residuals of the consistency plot in Fig.~\ref{fig:cloudfree_consistency} are centered around the horizontal black line at zero within the 1-10 $\mu$m wavelength range, we observe a slight offset above the zero line in the 10-16 $\mu$m range. This offset is not observed in retrievals performed exclusively on each individual dataset as shown in the Figs.~\ref{fig:consistency_onlyMIRI},~\ref{fig:consistency_onlyGemini} and~\ref{fig:consistency_onlyHST}. 
    Although not significant, this suggests a challenge in reconciling the near-infrared and MIRI spectra, which may stem from the absence of a more comprehensive forward model that accounts not only for bulk chemical processes but also for localized chemistry. One such instance is the deep atmospheric dynamics driven by fingering convection under dis-equilibrium chemistry, which can cause compositional gradients to impact the different regions of the spectrum differently, as discussed by \citet{2015ApJ...804L..17T}. Local effects become increasingly important while analyzing such long spectral wavelength ranges that provide deeper insights into larger parts of the atmosphere. Alternatively, the slight discrepancy could arise from an unknown process not accounted for in the forward models such as a missing opacity source deeper in the atmosphere \citep{Morley_2018, 2024ApJ...973...60B}. 
    Towards this effect, we performed tests including water clouds. However,they were not only not statistically preferred over the cloud-free retrievals, but they also did not account for this difference. Additionally, it could also be the models attempting to adjust for variations in the temperature continuum slopes, or from suspected near-infrared variability, estimated to be between 5\% and 30\% in the near-infrared \citet{leggett2016near}. 
    
    

    \subsection{Disequilibrium chemistry}
    \label{sec:disequilibrium_chem}

\begin{figure}[!t]
        \centering
        \includegraphics[width=\linewidth]{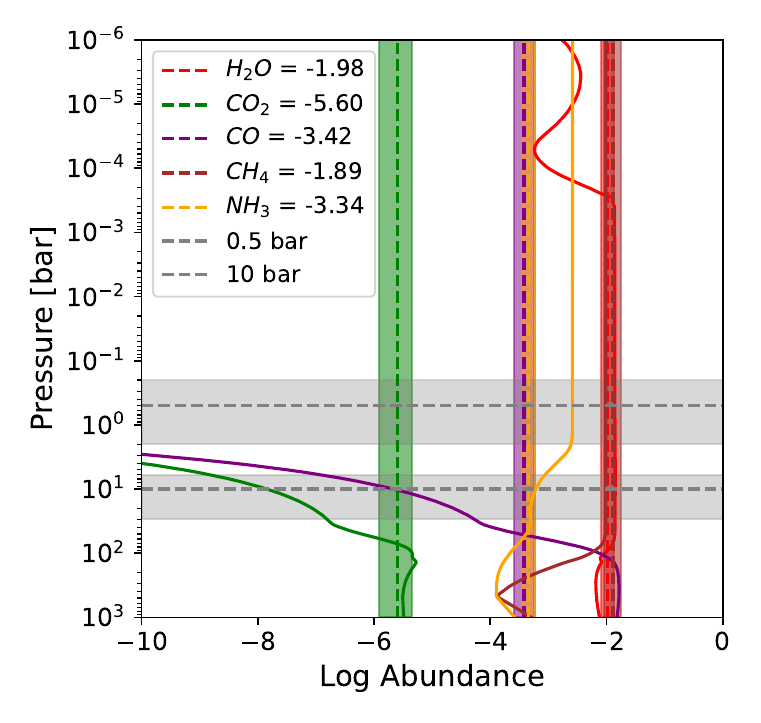}
        \caption{Chemical equilibrium abundances for a solar-composition atmosphere, calculated using the retrieved most probable $P$-$T$ profile. These are compared with the retrieved molecular abundances in mass fractions (in dashed lines, including 1$\sigma$ uncertainties as colored bars) for key opacity-contributing species: H$_2$O (red), CO$_2$ (green), CO (purple), CH$_4$ (brown), and NH$_3$ (orange). The grey regions are the estimated 1$\sigma$ of the emission contribution functions for the MIRI probed photosphere (at the top) and the near-infrared probed photosphere (at the bottom).}
        \label{fig:abundaces_all_cloudfree}
    \end{figure}

    Elemental abundances in substellar objects are key to understanding their evolutionary history, as they govern atmospheric opacity and cooling rates \citep{2001RvMP...73..719B}. These abundance patterns suggest either star-like formation by gravitational collapse or planet-like formation through gravitational instability, situating the sub-stellar brown dwarfs in the gap between higher-mass stars and lower-mass planets. Unlike stars, which directly reveal atomic abundances, the cooler atmospheres of brown dwarfs exhibit molecular abundances that can be used to determine elemental compositions. Additionally, molecular abundances not only reflect the atmospheric chemistry but also the dynamics at play within these atmospheres. In some instances, these molecular abundances are sensitive to equilibrium condensate rainout and vertical disequilibrium mixing \citep{2001RvMP...73..719B, 2007ApJS..168..140S}. In Fig.~\ref{fig:abundaces_all_cloudfree}, we compare the  retrieved molecular abundances, expressed as mass mixing ratios, for the key opacity-contributing species H$_2$O, CO$_2$, CO, CH$_4$, and NH$_3$ with the easyCHEM chemical equilibrium calculations for a solar composition atmosphere tabulated in \texttt{petitRADTRANS} \citep[see,][]{molliere2017observing, 2024arXiv241021364L}, in order to gain insights into the atmospheric dynamics of WISE 1738. 
    
    The retrievals constrain the abundances for CO and CO$_2$, even though their chemical equilibrium abundances are expected to be quenched at pressures lower than 1 bar. The fact that these species can be constrained to values higher than expected for chemical equilibrium using data from a higher photospheric region (illuminated by the MIRI spectrum), while remaining unconstrained in the near-infrared retrievals (unexpected under chemical equilibrium), provides clear evidence of disequilibrium mixing. This is explored in detail by \citet{2024ApJ...973...60B}, who investigate the over-abundance of CO$_2$ in brown dwarf atmospheres and suggest vertical mixing as a reason for this enhancement, despite an expectation to be quenched. These significant deviations suggest that additional processes, such as vertical mixing or chemical kinetics, need to be incorporated to fully explain the observed abundances with self-consistent models. 
    
    In contrast, the retrieved opacities
     of H$_2$O and CH$_4$ align well with equilibrium values consistent with previous studies \citep{1999ApJ...512..843B, 2007ApJS..168..140S}. However, the retrieved abundance of NH$_3$ is lower than its equilibrium value in the MIRI probed photosphere which is likely depleted due to quenching \citep{Zahnle_2014}, but is consistent with the near-infrared probed photosphere lower in the atmosphere. Consistency with equilibrium abundances seen in H$_2$O and CH$_4$ confirms a constant abundance profile for those molecules within the dis-equilibrium state of the atmosphere. 

    Additionally, we calculate a ratio of ${^{14}}$NH$_3$/${^{15}}$NH$_3$ which has recently been proposed as a new tracer for formation history \citep{barrado202315nh3}. Our measurement, which is non-robustly constrained with a 3$\sigma$ lower bound value of 275, largely overlaps with the range found for WISE 1828 ($670^{+390}_{-211}$) and the solar system. However, it also includes much lower estimates, aligning more closely with the estimate of $332^{+39}_{-43}$ for WISE 0855 reported by \citet{2024arXiv241010933K} and the ISM. Such broad values fail to identify core collapse as the formation pathway for brown dwarfs, as opposed to core accretion of exoplanets. Additional measurements of isotopologues in the atmosphere of WISE 1738 could help us understand the connections to formation scenarios further. 
    
    \subsection{Evolution of WISE 1738} 
        \label{sec:evolution}

    We estimate the effective temperature of WISE 1738. 
    This calculation relies on the bolometric fluxes over a broad wavelength range of 0.8 to 50 $\mu$m, derived from the retrieved posterior. Assuming a distance of 7.34 pc \citep{Martin_2018} and the retrieved radius posterior, the effective temperature is determined to be 402$^{+12}_{-9}$~K, which is in agreement with previous estimates in Table~\ref{table:summary_data_models}. Additionally, the $\log g$ is calculated using the retrieved values for mass and radius, yielding a result of 4.43$^{+0.26}_{-0.34}$.

    The evolutionary models presented by \citet{saumon2008evolution} and \citet{2021ApJ...920...85M} suggest that a brown dwarf with an effective temperature of approximately 400~K and a $\log g$ between 4.1 and 4.7~cm~s\(^{-2}\) would have a mass in the range of 6--20~\(M_J\), a radius between 0.97 and 1.1~\(R_J\), with a bolometric luminosity  $\log L/L\textsubscript{\(\odot\)}$ between $-6$ and $-7$. The retrieved mass of \(13^{+11}_{-7}\,M_J\), a radius of \(1.14\,R_J\), and bolometric luminosity of $-6.52^{+0.05}_{-0.04}$ align reasonably well with these theoretical predictions, supporting consistency between observed and modeled parameters. Further, we estimate an age spanning 1 to 4 Gyrs using the evolutionary model from \citet{2021ApJ...920...85M}, which is consistent with the rotation period of 6 hours measured for WISE~1738 \citep{leggett2016observed}, as brown dwarfs are expected to spin faster as they age \citep{2014prpl.conf..433B}. 
    

    \subsection{C/O and metallicty}
            \label{sec:CO_met}

    We determine the carbon-to-oxygen (C/O) ratio to be \(1.35^{+0.39}_{-0.31}\), and metallicity [M/H] to be \(0.34^{+0.12}_{-0.11}\), under the considered atmospheric model, assuming no oxygen sequestration in the atmosphere.
    However, the derived atmospheric abundances may not reflect the true chemical composition of the atmosphere because some chemical elements can be used up in cloud particles and/or are removed from the atmosphere due to rainout. Atmospheric oxygen is typically depleted by 20–30\% relative to intrinsic values due to sequestration in condensates, with the extent of depletion depending on the intrinsic metallicity and C/O ratio \citep{Line_2015}. This process increases the atmospheric C/O ratio while reducing the overall atmospheric metallicity, as oxygen is a dominant metal. Specifically, we account for enstatite and forsterite condensation \citep{1994Icar..110..117F} by adjusting the abundances of oxygen-bearing molecules by a factor of 1.3, following the approach in \cite{zalesky2019uniform}. This adjustment is equivalent to the removal of 3.28 oxygen atoms for every silicon atom \citep{1999ApJ...512..843B}. Incorporating this sequestration, we recalculate the C/O ratio and metallicity [M/H] as \(1.04^{+0.30}_{-0.24}\) and \(0.40^{+0.12}_{-0.10}\), respectively. These values lie within 2$\sigma$ and 4$\sigma$ of solar values respectively. 
     Such super-solar C/O ratios and metallicity have been observed in late T-dwarfs, as reported by \citet{Line_2017, zalesky2019uniform}. This ratio also aligns with the upper range of the local FGK population, which extends to a C/O ratio and metallicity upto values 1.4 and 0.6 respectively \citep{zalesky2019uniform, 2014AJ....148...54H}. 
     
     Brown dwarfs are thought to form via gravitational collapse that should result in near-solar C/O ratios and metallicity. 
     Interestingly, \citet{2013ApJ...779..178P} and more recently \citet{2023NatAs...7..805T, 2024Sci...384.1086A}, find the inner disk to be carbon-rich with molecules such as C$_{2}$H$_{2}$, HCN, C$_{6}$H$_{6}$, CO$_{2}$, HC$_{3}$N, C$_{2}$H$_{6}$, C$_{3}$H$_{4}$, C$_{4}$H$_{2}$, and CH$_{4}$ dominating the disk with little traces of H$_{2}$O (Arabhavi et al. 2025, submitted), suggesting that such values could be an artifact of formation. However, further understanding of oxygen sequestration processes and formation scenarios is necessary to explain such a high C/O ratio and metallicity. 

\section{Conclusion}
\label{sec:conclusion}

    Y dwarfs are among the coldest and least luminous brown dwarfs, offering unique opportunities to explore the thermal, chemical, and evolutionary properties of cool substellar objects. Their predominantly cloud-free atmospheres and spectra shaped by strong water vapor, methane and ammonia absorption, provide a large spectral dynamic range crucial for probing their pressure-temperature ($P$-$T$) profiles \citep[e.g.][]{1996Sci...272.1919M, Line_2015}. These objects serve as ideal laboratories for studying key atmospheric species such as H\(_2\)O, CH\(_4\), CO, and NH\(_3\) along with its isotopologue $^{15}$NH$_3$, which dominate the carbon and oxygen chemistry. This enables a precise determination of metallicity and carbon-to-oxygen (C/O) ratios, which are keys to link atmospheric compositions to evolutionary models. 
    
    While previous retrievals have primarily focused on near-infrared wavelengths, the inclusion of mid-infrared data from the Mid-Infrared Instrument of the James-Webb Space Telescope, allows us to probe higher in these atmospheres offering improved constraints on physical parameters like surface gravity, radius, mass, luminosity and unveiling dis-equilibrium chemistry by more accurately constraining chemical abundances. In this study, we build a comprehensive understanding of the Y dwarf WISE 1738 by retrieving its $P$-$T$ profile, chemical abundances (including an isotolpologue), comparing these findings to evolutionary models, and setting the stage for future investigations into complex atmospheric processes and cloud formation.
    
    We perform atmospheric retrievals using neural posterior estimation on the combined spectral data from WFC3+GNIRS+MIRI observations. To estimate their posterior distribution, we train a normalizing flow based on amortized variational inference algorithm, using the atmospheric emission models generated with the simulator \texttt{petitRADTRANS}. By combining retrievals over an extended wavelength range, we demonstrate a reduction in uncertainty in the retrieved $P$-$T$ profile and in some molecular abundances, compared to individual retrievals from each observation. We also caution against potential biases in spectral characterization that may arise from using narrower wavelength ranges, highlighting improved consistency between the estimated posterior and the observations when combined spectral data is used instead of individual spectral retrievals. This confirms trends that we have seen in short wavelength retrievals. 
    
    Additionally, we constrain the bulk physical properties of WISE 1738  including mass, radius, surface gravity, and luminosity, with higher accuracy, in agreement with theoretical predictions from evolutionary models. We also estimate the object's age to be between 1 and 4 Gyr which is consistent with the fast rotation period of 6 hrs. However, reconciling the near-infrared and mid-infrared regions proves challenging, suggesting the presence of an unknown process such as local chemistry, a missing opacity source, or intrinsic variability, that is not yet accounted for.

    In addition to the major opacity species such as CH$_4$ and H$_2$O, we also estimate the abundances of CO$_2$, CO, and the depleted NH$_3$, as well as its isotopologue $^{15}$NH$_3$ for the first time on this object. These results provide evidence of disequilibrium chemistry in WISE 1738's atmosphere due to vertical mixing, as they could not be constrained by the near-infrared data alone and equilibrium chemistry predicts their depletion below the near-infrared photosphere. The abundances of CH$_4$ and H$_2$O, however, are consistent with equilibrium chemistry. This result adds to the evidence of vertical mixing in brown dwarf atmospheres.      
 
\begin{acknowledgements}
      This project has received funding from the European Research Council (ERC) under the European Union’s Horizon 2020 research and innovation programme (grant agreements No 819155), and from the Wallonia-Brussels Federation (grant for Concerted Research Actions). 
      Computational resources have been provided by the Consortium des Équipements de Calcul Intensif (CÉCI), funded by the Fonds de la Recherche Scientifique de Belgique (F.R.S.-FNRS) under Grant No. 2.5020.11 and by the Walloon Region. 
      JPP acknowledges financial support from the UK Science and Technology Facilities Council, and the UK Space Agency. For the purpose of open access, the authors have applied a Creative Commons Attribution (CC BY) licence to the Author Accepted Manuscript version arising from this submission.”
\end{acknowledgements}

%
%
\bibliographystyle{aa} 

  


\begin{appendix} 

\section{Technical details on NPE}
\label{sec:NFarch}


      In this work, the normalizing flow is defined as a neural autoregressive flow \citep[NAF,][]{huang2018neural} implemented in the lampe package \footnote{\url{https://github.com/probabilists/lampe}}. It takes the model parameters and the ``context'' as inputs. Here, the context includes either spectral simulations from the training set, or the observation spectrum, depending on whether one is in the training or the inference phase. The flow outputs an estimate of the posterior distribution $p_{\phi}(\theta|x)$. The NAF is composed of several transformations parameterized by a neural network. Each transformation network is defined by a monotonic multilayer perceptron (MLP) and a signal network that conditions the MLP. The signal network is an autoregressive conditional function over the model parameters, taking these parameters and the context as input, and giving a signal vector that conditions the MLP as the output. The NAF is parameterized by three transformations, each defined by an MLP with five hidden layers of size \num{512}. These MLPs are conditioned on a signal output of length 16, and use Exponential Linear Unit (ELU) activation functions \citep{clevert2015fast}.  

     
     Before the context is input into the NAF to condition the transformations, it is compressed using an ``embedding'' network. This network embeds the spectrum such that important features are extracted instead of memorizing the training set, thus preventing overfitting. The embedding network is implemented as a ResidualMLP (or ResMLP), one for each region of the spectrum. The ResidualMLP networks consists of several linear blocks that decrease in size \citep{he2016deep}. The MIRI embedding network contains 10 residual blocks with dimensions 2 $\times$ \num{512} + 3 $\times$ \num{256} + 5 $\times$ \num{128}, whereas the WFC3+GNIRS embedding network contains 15 residual blocks with dimensions 3 $\times$ \num{512} + 5 $\times$ \num{256} + 7 $\times$ \num{128}, that are used to compress the length of the input features from (129+305)+1298, to a vector of feature length 8+64. Each of these embedding networks use ELU as their activation functions. The loss function \texttt{NPEloss} is used to optimize the training. 

     During training the optimization of the normalizing flow is carried out through a variant of stochastic gradient descent, namely AdamW \citep{loshchilov2018decoupled}. The initial learning rate used for training is $10^{-3}$, which is halved every time the average loss over the validation set does not improve for \num{32} continuous epochs, until it hits the value $10^{-8}$. This prevents overfitting \citep{zhang2021dive}. A weight decay of $10^{-2}$ is used for the training. The normalizing flow is trained for a total of 100 epochs, during which \num{1700} random batches of \num{1024} and \num{256} pairs each of $(\theta, f(\theta))$ are used from the training sets. The architectural hyper-parameters are adjusted on the validation data. 
        
      To arrive at the above architectures, we performed extensive hyper-parameter tuning on the flow, embedding network parameters and the loss function. For robustness, the twelve best configurations out of the 128 were chosen to perform all the retrievals conducted in this paper. The results from the best model among them are presented here. However, the variation of the results observed between these architectures are minimal. For the flow, we explored different numbers of transforms and hidden layer dimensions in the range of $[3,5]$ and $[256,512]$, respectively. For the embedding network, we tried different number of layers in the ResMLP in the range of $[10, 15]$ for each instrument, along with different embedding outputs in the range of $[4, 16]$ for the WFC3+GNIRS spectra.  We explored both the regular \texttt{NPEloss} and the \texttt{BalancedNPEloss} functions to penalize the overconfident posteriors \citep{delaunoy2023balancing}. We explored different values for the initial learning rate and minimum learning rate in the range of $[10^{-5}, 10^{-3}]$ and $[10^{-8}, 10^{-5}]$ respectively, and weight decay in the range of $[10^{-2}, 0]$. We analyzed the impact of different schedulers such as \texttt{ReduceLROnPlateau} and \texttt{CosineAnnealingLR} available in PyTorch, that allows dynamic learning rate reducing based on the validation loss, with patience rates between $[8,32]$. We tried batch sizes between $[2^8, 2^{11}]$ and the number of epochs between $[2^7, 2^{10}]$. 
    
      Similar to \citet{2023A&A...672A.147V}, amongst all the parameters that we tuned, the parameter weight decay between $[0, 10^{-2}]$ had the most significant impact on the training. We think this is because of the high variance of the input dataset, where some spectra are six orders of magnitude brighter than the rest. This leads to the skewing of the weights to very high values, which is compensated by weight decay to improve training performance. For more details, we refer to the source code of the experiments \footnote{\url{https://github.com/MalAstronomy/WISEJ1738}}.



\section{Diagnostics}
\label{sec:exp-diagnostics}

    The posterior density estimator, trained on simulated data is subjected to three diagnostic tests to evaluate their validity: the qualitative consistency plot, the coverage test, and the quantitative L-C2ST.

    \begin{figure*}[!t]
    \centering
    \includegraphics[width=0.9\textwidth]{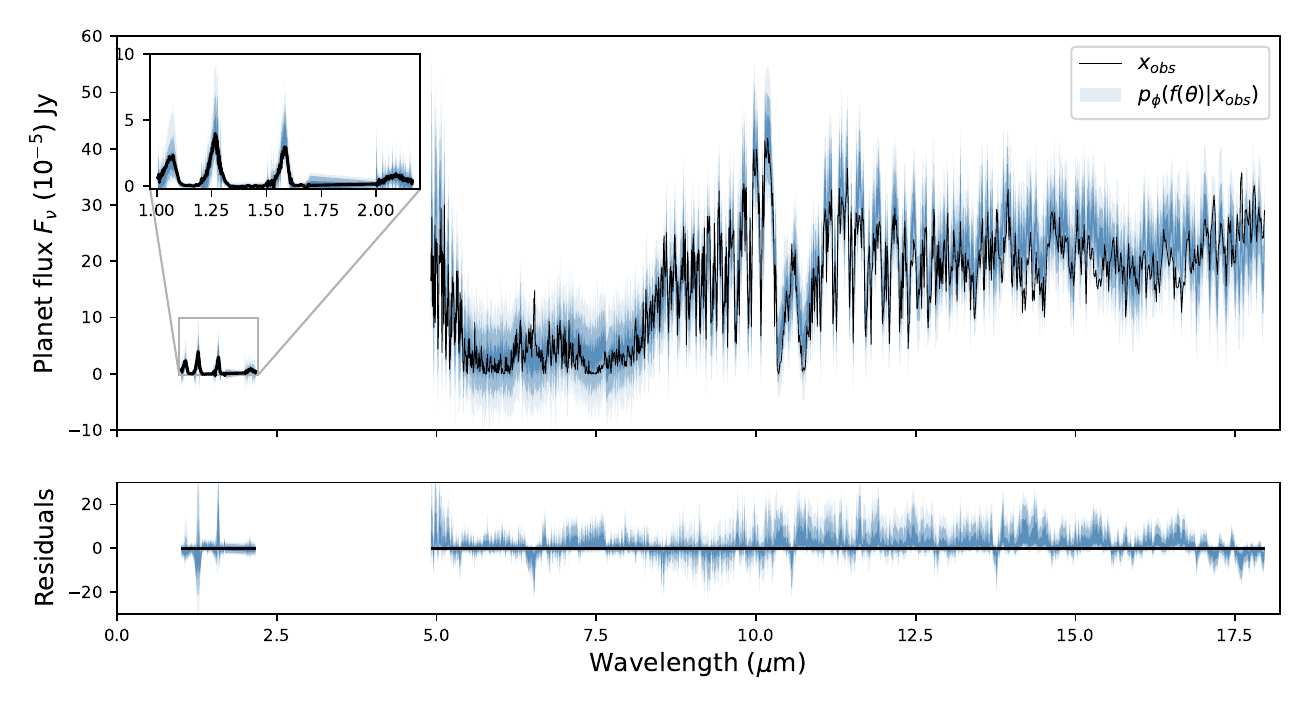}
    \caption{\textit{Top.} WFC3+GNIRS+MIRI consistency plot. The posterior predictive distribution $p(f(\theta)+ \epsilon |x_\text{obs})$ of noisy simulations spectra for the $99.7\%$, $95\%$ and $68.7\%$ quartiles (hues of blue), overlaid on top of the the WFC3+GNIRS+MIRI observation $x_\text{obs}$ (black line). {\textit{Bottom.} Residuals of the posterior predictive samples, normalized by the inflated standard deviation of the noise distribution for each spectral channel and a horizontal line at 0 for reference (in black).
    }}
    \label{fig:cloudfree_consistency}
    \end{figure*}


    \subsection{Consistency plot}
    \label{sec:consistency_check}

    A qualitative test to evaluate the consistency of the approximated posterior with the observation is by comparing the posterior predictive distribution  $p(x'| x)$ \citep{2023A&A...672A.147V} with it. This represents a distribution of the possible future observations $x'$, given the current observation $x$ and the model parameters $\theta$. It is calculated as $ p(x' \mid x) = \int p(x' \mid \theta) p(\theta \mid x) \, d\theta $
    where $p(x' \mid \theta)$ is the likelihood of the new data given the model parameters, and $p(\theta \mid x)$ is the estimated posterior distribution. The posterior predictive distribution is obtained by sampling parameters from the posterior \(\theta \sim p_\phi(\theta|x_\text{obs})\), and then computing their spectra \(f(\theta)+\epsilon\) using the simulator (which includes noise). One evaluates the quality of the fit by visually comparing the consistency of the posterior predictive distribution with the observation. 

    The cloud-free posterior predictive distribution  for the WFC3+GNIRS+MIRI spectra are displayed in Fig.~\ref{fig:cloudfree_consistency}. The results indicate that the posterior predictive distribution is well constrained, with the $x_\text{obs}$ centered well within the noise limit in most wavelengths with a small spread. We also plot the residuals normalized by the inflated error bars across all wavelengths.

\begin{figure*}[!t]
    \centering
    \includegraphics[width=0.9\textwidth]{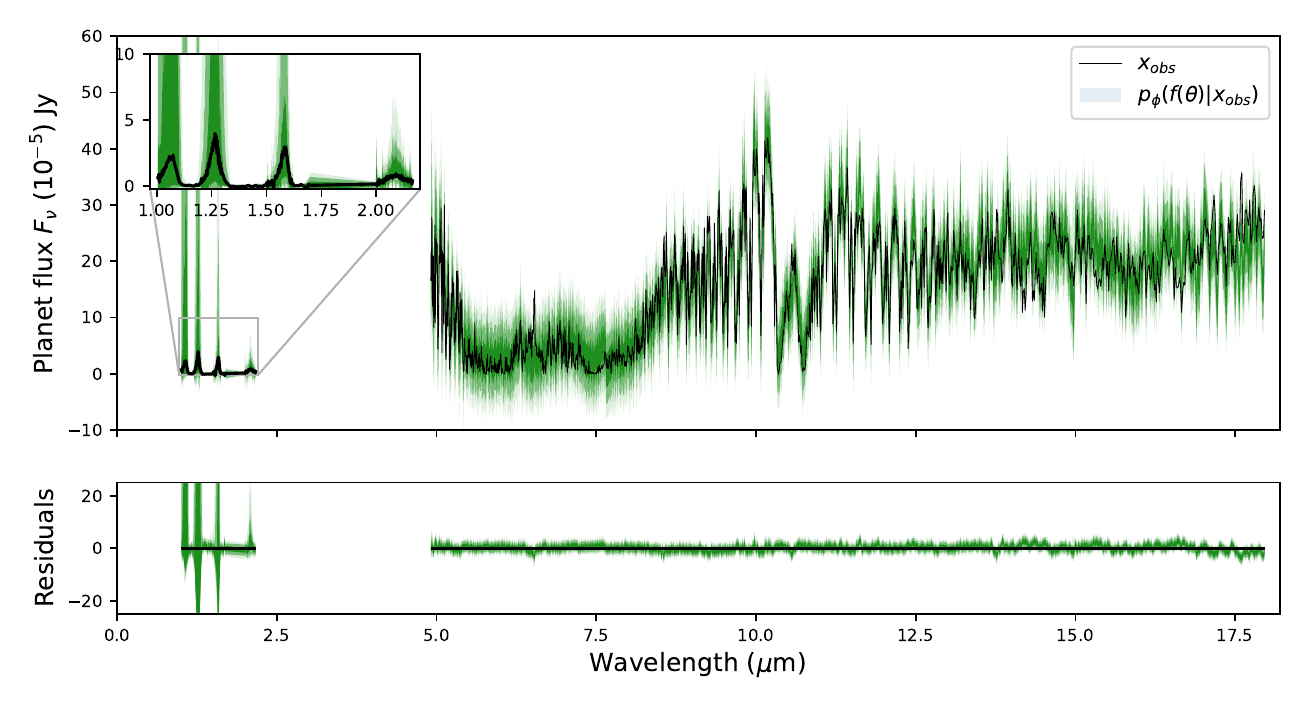}
    \caption{\textit{Top.} MIRI consistency plot. Posterior predictive distribution $p(f(\theta)+ \epsilon |x_\text{MIRI})$ of noisy simulations spectra for the $99.7\%$, $95\%$ and $68.7\%$ quartiles (hues of blue) obtained from the retrieval on MIRI data alone extended to near-infrared wavelengths, and overlaid on top of the the WFC3+GNIRS+MIRI observation $x_\text{obs}$ (black line). {\textit{Bottom.} Residuals of the posterior predictive samples, normalized by the inflated standard deviation of the noise distribution for each spectral channel and a horizontal line at 0 for reference (in black).
    } }
    \label{fig:consistency_onlyMIRI}
    \end{figure*}

\begin{figure*}[!t]
    \centering
    \includegraphics[width=0.9\textwidth]{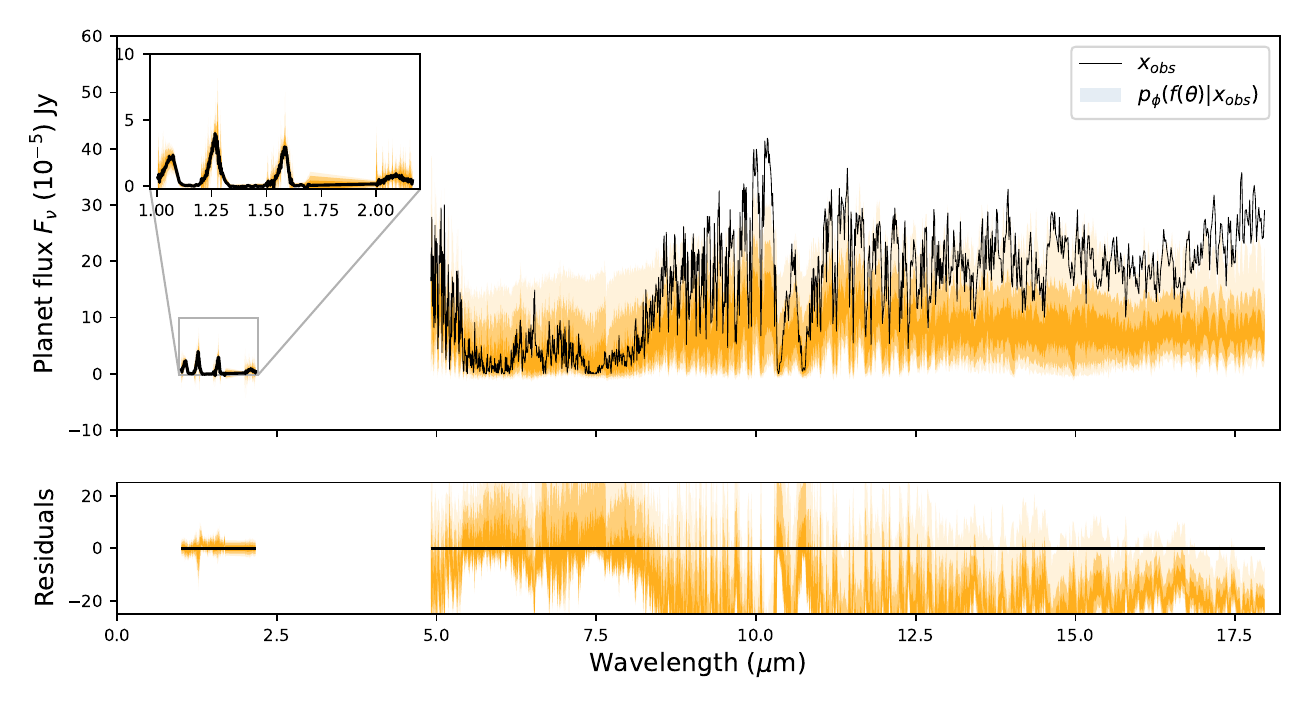}
    \caption{\textit{Top.} GNIRS consistency plot. Posterior predictive distribution $p(f(\theta)+ \epsilon |x_\text{GNIRS})$ of noisy simulations spectra for the $99.7\%$, $95\%$ and $68.7\%$ quartiles (hues of green) obtained from the retrieval on GNIRS data alone extended to mid-infrared wavelengths, and overlaid on top of the the WFC3+GNIRS+MIRI observation $x_\text{obs}$ (black line). {\textit{Bottom.} Residuals of the posterior predictive samples, normalized by the inflated standard deviation of the noise distribution for each spectral channel and a horizontal line at 0 for reference (in black).
    }}
    \label{fig:consistency_onlyGemini}
    \end{figure*}

    Similarly, in Figs.~\ref{fig:consistency_onlyMIRI} and \ref{fig:consistency_onlyGemini}, we compare the posterior predictive distributions for the independent posteriors obtained from separate retrievals on the MIRI and GNIRS spectra, each extended to the wavelength range covered by the WFC3+GNIRS+MIRI spectra. The inconsistencies observed in spectral regions not included in the original retrieval highlight potential biases in characterization when data availability is limited.
    
        \begin{figure*}[!t]
    \centering
    \includegraphics[width=0.9\textwidth]{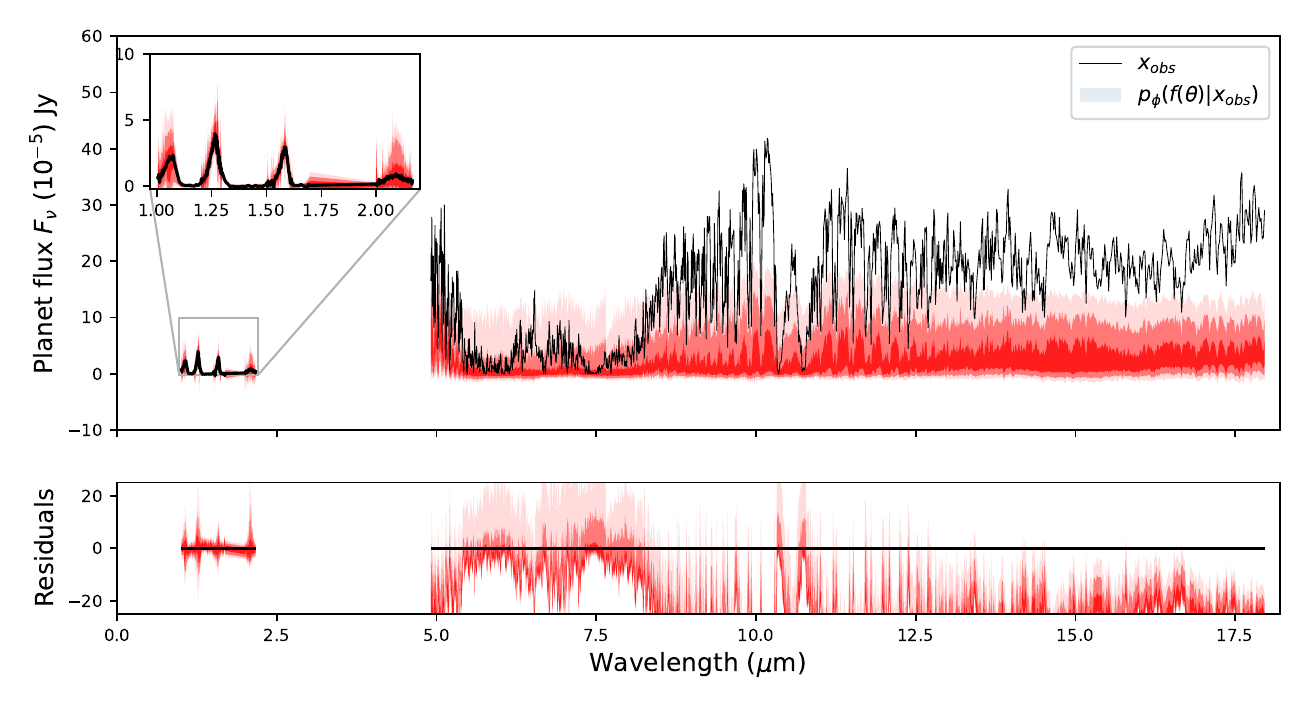}
    \caption{\textit{Top.} WFC3 consistency plot. Posterior predictive distribution $p(f(\theta)+ \epsilon |x_\text{WFC3})$ of noisy simulations spectra for the $99.7\%$, $95\%$ and $68.7\%$ quartiles (hues of red) obtained from the retrieval on WFC3 data alone extended to mid-infrared wavelengths, and overlaid on top of the the WFC3+GNIRS+MIRI observation $x_\text{obs}$ (black line). {\textit{Bottom.} Residuals of the posterior predictive samples, normalized by the inflated standard deviation of the noise distribution for each spectral channel and a horizontal line at 0 for reference (in black).
    }}
    \label{fig:consistency_onlyHST}
    \end{figure*}
   
    \subsection{Coverage}

    \begin{figure}[!t]
        \centering
        \includegraphics[width=\linewidth]{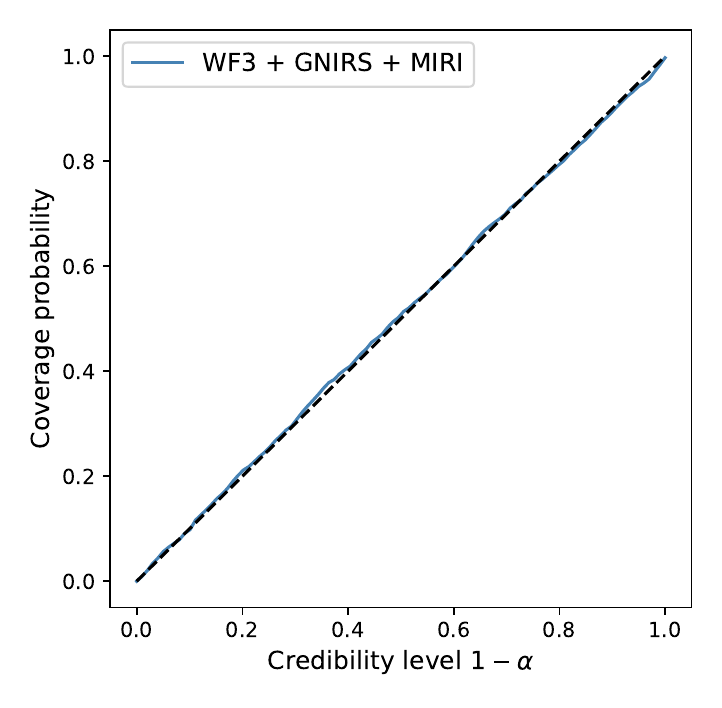}
        \caption{Coverage plots for cloud-free model posterior estimator. This plot suggests that the estimator is well calibrated. }
        \label{fig:coverage_cloudfree}
    \end{figure}
    
    Coverage is the probability of the true value of the model parameters appearing in a credible region of the estimated posterior distribution. It evaluates the consistency of the approximated posterior to assess whether it is under or over dispersed on average. 
      The coverage test is conducted by performing numerous retrievals on several simulated noisy spectra from the test set, and comparing their corresponding approximate posterior distributions with their nominal model parameter values. Based on the work by \citet{hermans2020likelihood}, the expected coverage probability of the $1-\alpha$ highest posterior density regions derived from the estimated posteriors $p_\phi(\theta|x)$ is defined as 
    \begin{equation}
    \label{eq:coverage}    
    \centering
    \mathbb{E}_{p(\theta,x)}\left[\mathds{1}\left(\theta \in \Theta_{p_\phi(\theta|x)}(1 - \alpha)\right)\right],
    \end{equation}
    where, $\mathds{1}(\cdot)$ is the indicator function, and the function $\Theta_{p_\phi(\theta | x)}(1 - \alpha)$ yields the $1 - \alpha$ highest posterior density region of $p_\phi(\theta | x)$. This is calculated as the percentage of times the nominal parameter values $\theta_{\text{test}}$ from the test set fall within a specified highest density region $(1-\alpha)$ of the corresponding estimated posterior $p_{\phi}(\theta_{\text{test}}|x_{\text{test}})$, where $x_{\text{test}} = f(\theta_{\text{test}})+\epsilon$. The computed coverage probability is plotted along credibility levels ranging between [0,1] on the x axis.  

    Figure~\ref{fig:coverage_cloudfree} plots the coverage probabilities for our NPE retrieval. Ideally if the posteriors are well calibrated, then $\theta_{\text{test}}$ should be contained in the $1-\alpha$ highest posterior density regions of the approximate posteriors  $p_{\phi}(\theta_{\text{test}}|x_{\text{test}})$  exactly $(1-\alpha)\%$ of the time. If the coverage probability is smaller than the credibility level $1 - \alpha$, then this indicates that the posterior estimator is overconfident. On the other hand, if the coverage probability is larger than the credibility level $1 - \alpha$, then this indicates that the posterior approximations are conservative. Here, we see that for the retrieval, the coverage plot almost aligns with the ideal case scenario (the dashed diagonal line). The probabilities align closely with a diagonal straight line, indicating well-calibrated posterior approximations.

    Given that this test requires several posterior estimations performed on the test set, it cannot be applied to non-amortized techniques, however it is straightforward in NPE. Here amortization is the ability to train the network once over one atmospheric model, and use it to perform retrievals over many observations almost quasi-instantaneously without having to start from scratch, which is a feature of NPE (and some other SBI algorithms). 


    \subsection{Local classifier two-sample test}

    Since coverage is a global validation method, it serves as a necessary but not sufficient condition for a valid inference algorithm. A coverage check that fails indicates that the inference is invalid, while passing coverage checks does not guarantee that the posterior estimation is accurate. This is because coverage deals with cumulative probability and does not account for local inconsistencies. This limitation motivates the use of a local validation procedure called the local classifier two-sample test \citep[L-C2ST,][]{linhart2024c2st}). L-C2ST allows for the local evaluation of a posterior estimator at any given observation. In case of an inconsistency, L-C2ST is also able to graphically show how to improve the estimator. 

    The test involves training a classifier to distinguish between samples drawn from the true joint distribution \( p(\theta, x) \) (class 0) and those drawn from the estimated posterior \( q(\theta \mid x)p(x) \) (class 1). For a normalizing flow, this translates to learning to differentiate samples from \( \mathcal{N}(0, \mathbf{I}_m)p(x) \) (class 0) and \( p(T_\phi^{-1}(\theta; x_o) \mid x_o) \) (class 1). The classifier is trained under the null hypothesis, which asserts that the two distributions are indistinguishable. Formally, the null hypothesis under the normalizing flow (NF) is expressed as:
    \[
    H_{NF,0}(x_{\rm obs}): \; p\left(T_\phi^{-1}(\theta; x_{\rm obs}) \mid x_{\rm obs}\right) = \mathcal{N}(0, \mathbf{I}_m)
    \]
    where, \( x_{\rm obs} \) is any observation over which one is evaluating the quality of the retrieval. 
    
    The classifier is also trained once on the observed data where the training set retains the relationship between variables to account for real-world variability. Finally, the classifier is evaluated on a single observation \( x_{\rm obs} \) based on metrics that rely on the L-C2ST statistic. 
    The statistic is the mean squared error (MSE) between 0.5 and the predicted probabilities from the classifier of being in class 0 over the dataset ($\theta_{\rm obs}, x_{\rm obs}$). 
    
    The first metric is the p-value, which is the proportion of the times the L-C2ST statistic under the null hypothesis is greater than the L-C2ST statistic at the observation \( x_{\rm obs} \). This is computed as the empirical mean over statistics obtained from several trials under the null hypothesis:
    \[
    p\text{-value} = \frac{1}{N} \sum_{i=1}^N \mathbb{I}\left( L_{\text{C2ST}}^{(i)} \geq L_{\text{C2ST}}(x_{\rm obs}) \right),
    \]
    where, \( L_{\text{C2ST}}(x_{\rm obs}) \) is the L-C2ST statistic at the observation \( x_{\rm obs} \), \( L_{\text{C2ST}}^{(i)} \) are the statistics under the null hypothesis, and \( \mathbb{I}(\cdot) \) is the indicator function. The distribution under the null hypothesis is called the T-distribution. If the posterior estimate is not close to the true posterior, the classifier will identify a significant difference between the two classes, resulting in higher values of the statistic and hence very small p-values. If this value is less than the assumed significance level, it indicates that the null hypothesis does not hold. 

    The binary classifier is implemented as an ensemble of 15 Multi-layer Perceptrons (MLP) from \texttt{scikit-learn}. The classifier is initialized with two hidden layers, each having a number of neurons equal to 10 times the number of input features (ndim = 26). The ReLU activation function is used in the hidden layers, and the Adam optimizer is employed to adjust the model’s weights. The training process runs for a maximum of 100 iterations. The classifier is trained on 50k samples from the training set. When trained, the classifier uses back-propagation to minimize the binary cross-entropy loss function, adapting its weights to improve predictions of one of two possible outcomes (e.g., 0 or 1) based on the input data.
    
\begin{figure*}[!t]
        \centering
        \includegraphics[width=\textwidth]{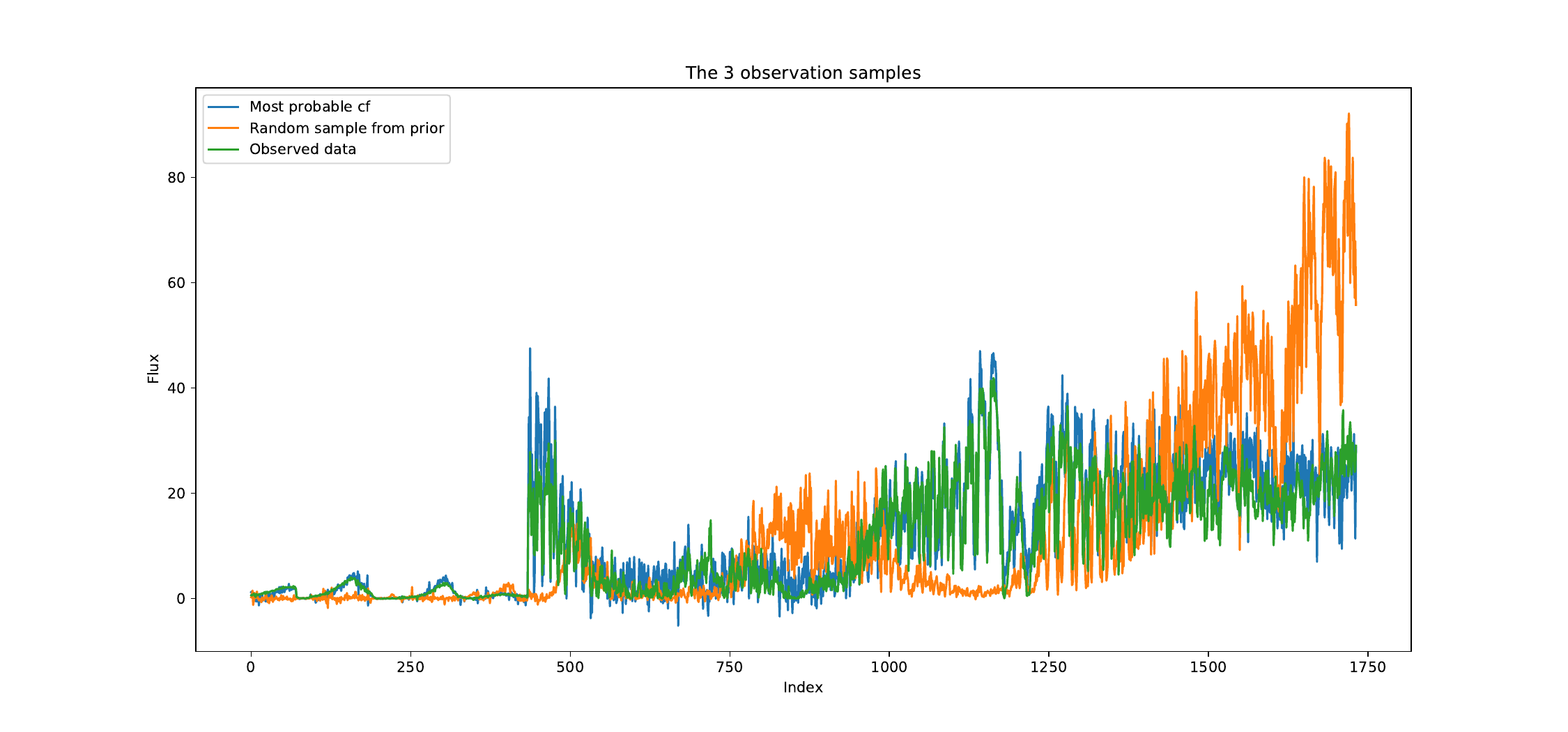}
        \caption{The three observations for which we evaluate the estimated posterior.}
        \label{fig:lc2st_obs}
    \end{figure*}

\begin{figure*}[!t]
        \centering
        \includegraphics[width=\textwidth]{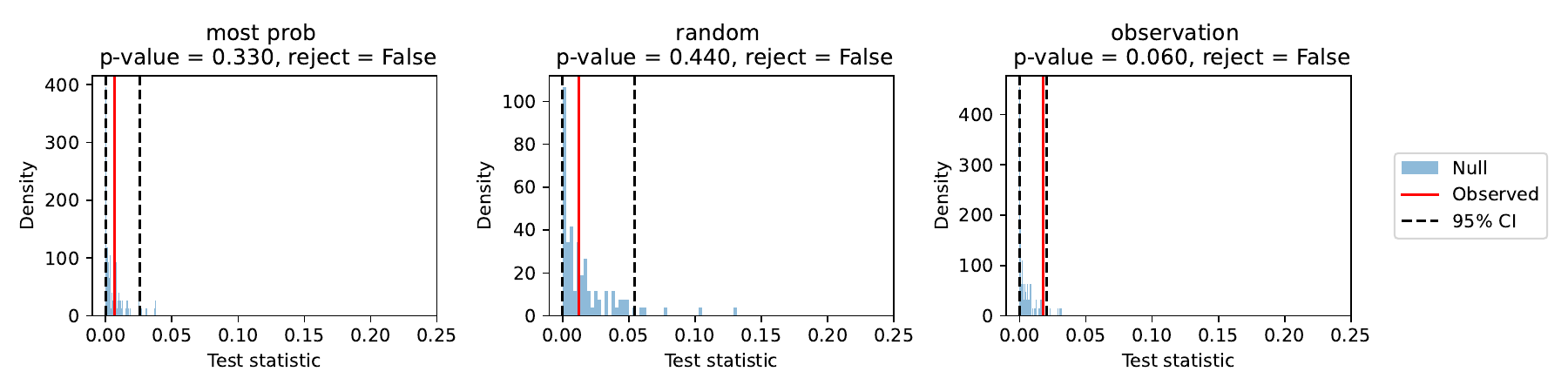}
        \caption{T distribution plot. The p-values are computed as the proportion of the times the L-C2ST statistic under the null hypothesis is greater than the L-C2ST statistic at the observation \( x_{\rm obs} \).  }
        \label{fig:Tdist}
    \end{figure*}

    We calculate the p-values for: the estimated posterior for WISE 1738's spectrum (Observation 3), the most probable simulated observation (Observation 1), and a random sample from the prior (Observation 2). This is illustrated in Fig.~\ref{fig:lc2st_obs}. 
    In each case, the T-distribution shown in Fig.~\ref{fig:Tdist} indicates a p-value well within the rejection threshold of 0.05, suggesting that the null hypothesis cannot be rejected. This implies that the estimated posterior for these observations is close to the true posterior. 

\begin{figure*}[!t]
        \centering        
        \includegraphics[width=\textwidth]{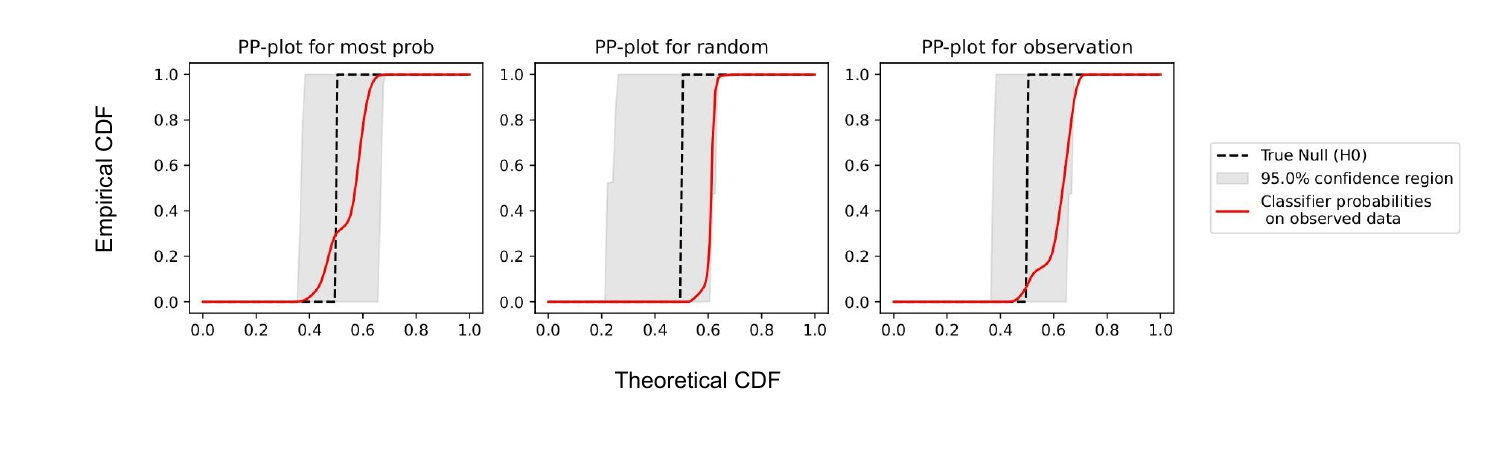}
        \caption{pp-plot. Cumulative distribution function (CDF) for the three posterior estimates. }
        \label{fig:pp_plot}
    \end{figure*}

    Next, we present the pp-plots for these observations (see Fig.~\ref{fig:pp_plot}). The pp-plot, a variation of the coverage plot, helps assess the overall trend of bias or the potential over- or under-confidence of the estimated posterior. For the estimated posterior of WISE 1738's spectrum and observation 3, the red curve is entirely within the gray confidence region, suggesting that it is neither significantly over-dispersed nor under-dispersed. However, there is a slight rightward bias (a small deviation from the dashed vertical line), indicating that some estimated parameters may be slightly higher than the nominal value. In contrast, Observation 1 shows less bias and is valid. 

\end{appendix}


\end{document}